\documentclass{aa}

\usepackage{natbib}
\bibpunct{(}{)}{;}{a}{}{,}

\usepackage[T1]{fontenc}
\usepackage[latin1]{inputenc}

\usepackage{amsmath}

\usepackage{graphicx}
\graphicspath{{./}{postscript/}{pdf/}}

\newcommand{\diff}[2][]{\,\mathrm{d}#2^{{#1}}}
\newcommand{\der}[3][]{\frac{\mathrm{d}^{{#1}}\,{#2}}{\mathrm{d}{#3}^{{#1}}}}

\newcommand{\pder}[3][]{\frac{\partial^{{#1}}\,{#2}}{\partial {#3}^{{#1}}}}

\newcommand{\Gz}{\ensuremath{{\cal G}_z}}

\newcommand{\Inu}{\ensuremath{I_\nu}}
\newcommand{\Jnu}{\ensuremath{J_\nu}}
\newcommand{\Hnu}{\ensuremath{H_\nu}}
\newcommand{\Knu}{\ensuremath{K_\nu}}
\newcommand{\Bnu}{\ensuremath{B_\nu}}
\newcommand{\Snu}{\ensuremath{S_\nu}}
\newcommand{\kapnu}{\ensuremath{\kappa_\nu}}
\newcommand{\chinu}{\ensuremath{\chi_\nu}}
\newcommand{\signu}{\ensuremath{\sigma_\nu}}
\newcommand{\fK}{\ensuremath{f_K}}
\newcommand{\fH}{\ensuremath{f_H}}

\newcommand{\sigmab}{\ensuremath{\sigma_B}}
\newcommand{\sigmaj}{\ensuremath{\sigma_J}}

\newcommand{\kappab}{\ensuremath{\kappa_B}}
\newcommand{\kappaj}{\ensuremath{\kappa_J}}
\newcommand{\kappah}{\ensuremath{\kappa_H}}
\newcommand{\kappar}{\ensuremath{\kappa_\mathrm{R}}}
\newcommand{\chij}{\ensuremath{\chi_J}}
\newcommand{\chih}{\ensuremath{\chi_H}}

\newcommand{\taus}{\ensuremath{\tau^\mathrm{s}}}
\newcommand{\Is}{\ensuremath{I^\mathrm{s}}}
\newcommand{\Js}{\ensuremath{J^\mathrm{s}}}
\newcommand{\Hs}{\ensuremath{H^\mathrm{s}}}
\newcommand{\Ks}{\ensuremath{K^\mathrm{s}}}
\newcommand{\Inus}{\ensuremath{\Inu^\mathrm{s}}}
\newcommand{\Jnus}{\ensuremath{\Jnu^\mathrm{s}}}
\newcommand{\Hnus}{\ensuremath{\Hnu^\mathrm{s}}}

\newcommand{\chihs}{\ensuremath{\chih^\mathrm{s}}}

\newcommand{\kappajs}{\ensuremath{\kappaj^\mathrm{s}}}

\newcommand{\omegas}{\ensuremath{\omega^\mathrm{s}}}

\newcommand{\tauz}{\ensuremath{\tau^0}}
\newcommand{\Iz}{\ensuremath{I^0}}
\newcommand{\Jz}{\ensuremath{J^0}}
\newcommand{\Hz}{\ensuremath{H^0}}
\newcommand{\Kz}{\ensuremath{K^0}}
\newcommand{\Inuz}{\ensuremath{I_\nu^0}}
\newcommand{\Jnuz}{\ensuremath{J_\nu^0}}
\newcommand{\Hnuz}{\ensuremath{H_\nu^0}}
\newcommand{\chijz}{\ensuremath{\chij^0}}
\newcommand{\chihz}{\ensuremath{\chih^0}}
\newcommand{\sigmajz}{\ensuremath{\sigmaj^0}}
\newcommand{\kappajz}{\ensuremath{\kappaj^0}}

\newcommand{\omegaz}{\ensuremath{\omega^0}}
\newcommand{\muz}{\ensuremath{\mu^0}}

\newcommand{\tauh}{\ensuremath{\tau_H}}
\newcommand{\dtauh}{\ensuremath{\Delta\tauh}}

\newcommand{\thetav}{\ensuremath{\theta_\mathrm{V}}}
\newcommand{\thetaa}{\ensuremath{\theta_\mathrm{A}}}
\newcommand{\dtauhv}{\ensuremath{(\dtauh)_\mathrm{V}}}
\newcommand{\dtauha}{\ensuremath{(\dtauh)_\mathrm{A}}}

\newcommand{\uv}{\ensuremath{u_\mathrm{V}}}
\newcommand{\ua}{\ensuremath{u_\mathrm{A}}}
\newcommand{\Uv}{\ensuremath{U_\mathrm{V}}}
\newcommand{\Ua}{\ensuremath{U_\mathrm{A}}}
\newcommand{\Teff}{\ensuremath{T_\mathrm{eff}}}
\newcommand{\Teffa}{\ensuremath{(\Teff)_\mathrm{A}}}

\newcommand{\Ta}{\ensuremath{{T_\mathrm{A}}}}
\newcommand{\mdot}{\ensuremath{{\dot M}}}
\newcommand{\nuvisc}{\ensuremath{\nu_{\rm visc}}}

\def\msyr{\ensuremath{M_\odot/\mbox{yr}}}
\newcommand{\latin}[1]{{\itshape #1}}

\begin{document}

\title{The vertical structure of T~Tauri accretion disks\\
  II. Physical conditions in the disk}
\author{Fabien Malbet\inst{1} \and Régis Lachaume\inst{1} \and Jean-Louis 
Monin\inst{1,2}}
\offprints{F.~Malbet,\\ \email{Fabien.Malbet@obs.ujf-grenoble.fr}}
\institute{Laboratoire d'Astrophysique UMR UJF-CNRS 5571, Observatoire de
  Grenoble, Université Joseph Fourier, BP 53, F-38041 Grenoble cedex 9,
  France\and
Institut Universitaire de France}
\date{18 May 2001 / 11 September 2001}
\authorrunning{Malbet, Lachaume \& Monin}
\titlerunning{The vertical structure of T~Tauri disks. II. Physical
  conditions in the disk}

\abstract{
  We present a self-consistent analytical model for the computation of the
  physical conditions in a steady quasi-Keplerian accretion disk.  The
  method, based on the thin disk approximation, considers the disk as
  concentric cylinders in which we treat the vertical transfer as in a
  plane-parallel medium. The formalism generalizes a work by \citet{hub90},
  linking the disk temperature distribution to the local energy dissipation
  and leads to analytical formulae for the temperature distribution which
  help to understand the behaviour of the radiation propagated inside the
  disks.  One of the main features of our new model is that it can take
  into account many heating sources.  We apply the method first to two
  sources: viscous dissipation and stellar irradiation.  We show that other
  heating sources like horizontal transfer or irradiation from the ambiant
  medium can also be taken into account. Using the analytical formulation
  in the case of a modified Shakura \& Sunyaev radial distribution that
  allow the accretion rate to be partly self-similar in the inner region,
  and, for an $\alpha$ and $\beta$ prescription of the viscosity, we obtain
  two-dimensional maps of the temperature, pressure and density in the
  close environment of low mass young stars.  We use these maps to derive
  the observational properties of the disks such as spectral energy
  distributions, high resolution spatial images or visibilities in order to
  underline their different behaviours under different input models.
}

\maketitle


\section{Introduction}
\label{sec:intro}

Since the initial models of viscous accretion disks of \citet{ss73} and
\citet{lbp74}, many efforts have focused on the description of the close
environment of T~Tauri stars (hereafter TTS).  In order to reproduce high
angular resolution images that became available soon after 1990, various
authors have published models of scattered light in large circumstellar
structures \citep{laz90,whi92}.  In the last five years, more precise
images of edge-on disks have been obtained by the \emph{Hubble Space
  Telescope} (HST) and infrared adaptive optics \citep{bur96,sta98,mb00},
allowing us to refine the models and the physical parameters used to
describe the disks.  Most of the time, when interpreting these images, the
authors use \latin{ad-hoc} power laws to extract the physical conditions in
the disks, like surface density or the disk scale height.  Indeed, these
images show that the large-scale flaring disks have a complex vertical
structure that must be taken into account in the interpretation of their
flat SED.

Up to now, we have access to only a very few constraints on the very
central parts of circumstellar disks, since current instruments do not
allow us to directly observe within 10 AU of the central object.  However,
the physics of the inner part of the disks is not expected to be completely
different from that in their outer parts, but our knowledge of these
central regions of circumstellar disks is restricted to scarce
interferometric data and indirect measurements of the inner phenomena via
the spectral energy distribution (hereafter SED), the high energy UV and X-ray
emission of disks or the measurement of magnetic fields.

The first models used quasi-Keplerian steady accretion disks, expected to
be geometrically thin for reasonable values of the accretion rate.  They
describe the disk as infinitely flat, predicting that the emergent flux has
a spectral energy distribution proportional to the power -4/3 of the
wavelength in the infrared.  However, \citet{rz87} showed that most TTS
present flatter spectral energy distributions, \latin{i.e.} decreasing less
steeply or even remaining constant over a large domain in the infrared.
This discrepancy led \citet{asl88} to assume that the radial temperature
distribution in T~Tauri disks is also flatter than in standard accretion
disks: if one assumes that the disk thickness flares then the disk
intercepts more light from the star at large distances and leads to a
variety of SED slopes.

Modeling of the vertical structure of the disks and the associated
radiative transfer has been developed by \citet{bl94} focusing their
attention on the vertical structure of disks in FU Orionis systems where
they explained a possible origin of outbursts.  \citet{cg97,cg99} used a
semi-analytical model of the circumstellar structure to explain the SEDs of
different types of young stars.  Recently, \citet{dal98,dal99} numerically
resolved a complete set of equations to compute the spatial distribution of
the temperature and the density that can be constrained by the observed
SEDs and two-dimensional images.

Up to now, the models have included various approximations~: (i) the
magnetic field is ignored; (ii) the viscosity is assumed to depend only on
local conditions and (iii) the viscosity is expressed using a simple
\latin{ad-hoc} law.  On the other hand, when authors take into account the
magnetic field \citep{shu94,fer95,cas00a,cas00b}, they assume that these
magnetic fields do not affect the overall disk structure.  As another
example, \citet{ter98a} and \citet{ter98b} showed that gravitational
instabilities in the disks may produce waves dissipating viscous energy far
from where they are created.  Integrating all these effects in a unique
model appears to be a difficult task.

Our goal is to develop a radiative model simple enough that it could later
on include as many detailed physical processes of the energy transport in
the disks as possible, and / or the interaction between the circumstellar
matter and the magnetic field.  In this paper, we present a new method to
process the radiative transfer in circumstellar disks, following the
pioneering work by \citet{hub90} and an initial application in the case of
TTS disks by \citet{mb91,mal92}.  Our results are not fundamentally
different from previous models but thanks to the analytical expression of
the vertical temperature distribution that we develop, we are able to
identify the actual origin of the temperature rise in the upper part of the
disk photosphere \citep{mb91}.  We propose that this formalism can be the
starting point for more elaborate models, including magnetic field or
viscous transport.  As a first step, our new self-consistent analytical
model can be used to constrain the physical environment of young stars
using images from HST or ground-based large telescopes, as well as
millimeter radio-interferometer maps and forthcoming data from large
optical interferometers, and especially the ESO VLTI project \citep{mal00}.

Section \ref{sec:model} presents the analytical model derived from hydrostatic
and  radiative equilibrium, as first described by \citet{hub90}, and  extended
here to the illumination by  an external source. Section \ref{sec:ttsdisks}
describes the application of this model to the specific case of the
circumstellar environment of young stellar objects, especially in the case
where the accretion rate is not constant at all radii. To illustrate the
model and its application, we present and discuss in Sect.~\ref{sec:results}
some results in the case where the heating by the central source can be
neglected and we focus our attention on the production of observable quantities
like SEDs, images, and interferometric data including visibility amplitudes and
closure phases.


\section{Analytical model}
\label{sec:model}

In this section, we detail the analytical model for the calculation of the
spatial distribution of quantities like the temperature or the density.  We use
the notations of \citet{hub90} as much as possible, even if we present the
formalism in a slightly different way. The first 3 subsections
(\ref{sec:model.init}, \ref{sec:model.hydro} and \ref{sec:radiat}) are short
reminders of the main equations of the problem. The last one
(\ref{sec:model.external}) is the generalization of this formalism to external
sources of radiation.

The principle for computing the vertical structure of T Tauri disks is to
consider the disk as a set of concentric cylinders of infinitesimal width.
Each cylinder is itself regarded as a plane-parallel atmosphere of mixed dust
and gas layers.  The basic assumption is that the radial radiative transfer of
energy is smaller than the vertical transfer, which is true only in the thin
disk approximation \citep{pri81}.  However, we can still take into account the
horizontal transfer as a perturbation for disks of moderate flaring (see in
Sect.~\ref{sec:htr}).  We solve the coupled equations of hydrostatic
equilibrium and the radiative transfer in each cylinder in order to compute the
temperature and density distributions along the $z$-axis at a given radius $r$.
In the following sections, when we focus our attention on the vertical
structure of the cylinder of radius $r$, we drop the $r$ coordinate unless
otherwise specified.  We also use the mass column coordinate $m$ in place of
the height coordinate $z$:

\begin{equation}
   m(z) = \int_{z}^{\infty} \rho(z)\diff{z},
\end{equation}
where $M = m(0) = \Sigma/2$ is half the surface density of the disk at
radius $r$.

\subsection{Initialization stage from the radial structure}
\label{sec:model.init}

For the initial step, we assume a standard geometrically thin disk with a
uniform vertical structure at each radius.  To fully solve the radial
structure, we need:
\begin{itemize}
   \item the mass per area unit $\Sigma(r)$
   \item the energy dissipation per area unit $F(r)$
   \item the opacity law $\chi(T, \rho)$ of the disk material
   \item the equation of state of the material, $\rho = {\cal S}(P, T)$
   \item the local vertical acceleration field $\Gz(r, z)$, generally
   inferred from the gravity
\end{itemize}
where the temperature, density, scale height, pressure and optical thickness
laws, respectively $T_{\rm D}$, $P_{\rm D}$, $\rho_{\rm D}$,
$h_{\rm D}$ and $\tau_{\rm D}$, are given by the coupled equations:
\begin{equation}
  \label{eq:rsT}
  \begin{split}
    T_{\rm D}^4 (r)  &= \frac{F(r)/2} {1-{\rm e}^{-\tau_{\rm D}(r)}} \\
    P_{\rm D}(r)     &= -\Gz\big(r, h_{\rm D}(r)\big) \, \Sigma(r)\\
    \rho_{\rm D}(r)  &= {\cal S}\big(P_{\rm D}(r), T_{\rm D}(r)\big)\\
    h_{\rm D}(r)     &= \Sigma(r)/\rho_{\rm D}(r)\\
    \tau_{\rm D}(r)  &= \Sigma(r) \, \chi\big(T_{\rm D}(r), \rho_{\rm
      D}(r)\big)\\ 
  \end{split}
\end{equation}

In the optically thin case, an iterative computation is required.  One should
notice that the mass, energy dissipation and acceleration field distributions
do not need to be independent of the temperature or density laws; in this case
   $\Sigma(r)$, $F(r)$ and $\Gz(r, z)$ are refined at each iteration.

A first refinement is to replace the uniform distribution $\rho_{\rm D}(r)$ by
the isothermal density distribution. The resulting radial structure is then the
structure given by the so-called \emph{standard} model \citep{ss73,lbp74}.  

Another improvement consists of replacing the effective temperature by the
temperature of the equatorial layer located at $\tau_{\rm D}/2$. As in a
stellar atmosphere, we substitute for the first equation of the set of equations
(\ref{eq:rsT}) the following equation:
\begin{equation}
T_{\rm D}^4 (r)  = 
      \frac{ 3/4 \, F(r) \, (\tau_{\rm D}(r) + 2/3) } { 
         1- {\rm e}^{-\tau_{\rm D}(r)} } \label{eq:rsT2}
\end{equation}


\subsection{Hydrostatic equilibrium --- Density distribution}
\label{sec:model.hydro}

For a given vertical temperature law, we compute the density distribution from
the pressure given by the hydrostatic equilibrium.  As the disk is dominated by
gravitation of the central star, the partial derivative of $\Gz$ is a constant
in the approximation $z \ll r$, and we use a differentiated hydrostatic
equilibrium equation for convenience:

\begin{equation}
\label{eq:hydro}
\der[2]{P}{m} = 
   - \frac{1}{{\cal S}(P, T)} \pder{\Gz}{z}
\end{equation}

The second order equation Eq.~(\ref{eq:hydro}) is solved with two following 
boundary conditions:
\begin{enumerate}
   \item in the disk mid-plane: the first derivative of $P$ is zero by symmetry.
   \item in the outer boundary: $P$ tends to a constant low value, the pressure
   in the interstellar medium (hereafter ISM). The choice of this limit has 
   almost no influence on the disk structure.
\end{enumerate}

\subsection{Radiative equilibrium --- Temperature distribution}
\label{sec:radiat}

To complete the calculations of the distributions of the physical quantities,
we have to connect the local temperature with the heating sources, remote or
local.  This is the goal of the radiative transfer described in this section.

Theoretically speaking, the radiative transfer equations can be applied to any
kind of radiation.  However, common and well-known approximations make them
easier to solve provided the radiation is thermalized or isotropic.  For
instance we can approximate the flux-weighted mean opacity by the Rosseland
opacity if the radiation is thermalized with the medium.  The Eddington closure
relations apply well for isotropic radiation.

We therefore  treat separately radiation emitted (from internal heating sources
or from external light absorbed by the disk) which is mainly isotropic and
thermalized, and the radiation coming from external sources.  The latter
can also be split into two terms, the attenuated radiation coming from
its external source in a certain known direction, and the radiation that has
been scattered one or several times which should be mainly isotropic. In the
first part of this section, following Hubeny we neglect the radiation by
external sources. Later in Sect.~\ref{sec:model.external}, we will take into
account correctly the influence of external radiations.

\subsubsection{Energy dissipation in each layer}

We first express the temperature as a function of the local production of
energy $u$, such as viscous heating, stellar irradiation, disk backwarming,
heating by a UV or X radiation field, or the contribution of the energy
horizontal transfer.  Following the usual notations \citep{mih78}, we link $u$
to the source function $S$ and the diffuse radiation intensity $I$:

\begin{gather}
   4\pi \int_0^\infty \chi_\nu(m) \left[ \Snu(m)-\Jnu(m) \right] \diff\nu
      = \rho(m) u(m)
   \label{eq:radeq}\\
   {\rm with} 
   \quad \Jnu(m) = \int_{-1}^{1}\Inu(m, \mu) \diff\mu
\end{gather}

The conservation of the radiative energy means that in an atmosphere layer, the
radiated energy, \latin{i.e.} the emission of the layer, minus the radiation
coming from the neighbouring layers $\Jnu(z)$, is exactly equal to the
dissipated energy in the layer:
\begin{equation}
\Snu =   \frac \kapnu\chinu \Bnu 
        + \left( 1 - \frac \kapnu\chinu \right) \Jnu,
\end{equation}

Integrating over frequencies, one derives from Eq.~(\ref{eq:radeq}):
\begin{equation}
   \kappab B = \kappaj J + \frac{u}{4\pi}
         \label{eq:KB(m)}
\end{equation}
where $\kappab$ and $\kappaj$ are respectively the Planck absorption mean and 
the {\it usual} absorption mean, defined by:
\begin{align}
   \kappab &=\int_{0}^{\infty}  (\kapnu / \rho) \Bnu \, {\rm d}\nu / B\\
   \kappaj &=\int_{0}^{\infty}  (\kapnu / \rho) \Jnu \, {\rm d}\nu / J
\end{align}

Since $\pi B = \sigma_B T^4$, the vertical temperature distribution is, by 
using Eq.~(\ref{eq:KB(m)}):
\begin{equation}
T^4 = \frac{\pi}{\sigmab}\,\frac \kappaj\kappab
         \left[ J + \frac{u}{4\pi\kappaj} \right]
\label{eq:Tfuncu}
\end{equation}
In other words, the temperature in a layer is the sum of two contributions: (i)
a term proportional to the radiative intensity $J$ transferred by adjacent
layers and (ii) a local dissipation term proportional to $u$.

\subsubsection{Radiative transfer of the diffuse intensity}
\label{sec:model.radiat}

We solve the transfer of the diffuse intensity $I$ to determine $J$. The 
equation of transfer and its two first moments are:
\begin{align}
   \mu\pder{\Inu}{z}(z,\mu) &=  \chinu(z) \left[ \Snu(z)-\Inu(z,\mu) \right]\\
      \der {\Hnu}{z}(z)     &=  \chinu(z) \left[ \Snu(z)-\Jnu(z)     \right], 
                                                \label{eq:1stmom}\\
      \der {\Knu}{z}(z)     &= -\chinu(z) H_\nu(z)
                                                \label{eq:2ndmom}
\end{align}
where $\Jnu$, $\Hnu$ and $\Knu$ are respectively the zeroth, first and
second moments of $\Inu$.

By integrating over frequencies one gets:
\begin{align}
\der{H}{m} &= -\frac{u}{4\pi} \label{eq:diffH}\\
\der{K}{m} &=  \chih H        \label{eq:diffK}
\end{align}
In this equation, the flux-averaged opacity $\chi_H$ is defined by:
\begin{equation}
\chih = \int_{0}^{\infty}  (\chinu/\rho) \Hnu \diff\nu / H,
\end{equation}

When integrating the first moment equation Eq.~(\ref{eq:diffH}) over $m$, one 
gets:
\begin{align}
                H(m)      &= H(0)   - U M \theta(m) / 4 \pi\\
{\rm with}\quad U M       &= \int_0^M u(\zeta) \diff{\zeta}\\
{\rm and} \quad \theta(m) &= \frac{1}{UM} \; \int_0^m u(\zeta) \diff{\zeta}
   \label{eq:theta}
\end{align}
where $U$ is the total energy dissipation in the upper atmosphere of the disk
per surface unit, and $\theta$ the distribution function of the energy
dissipation in this atmosphere.

The symmetry conditions in the equatorial plane for the  flux $H(m)$ implies
that $H(M)=0$, that is $\theta(M) = 1$. We therefore get a relation between
$H(0)$ and the energy flux dissipated in all atmospheric layers $U$:
\begin{equation}
H(0) = U M / 4 \pi
\label{eq:U}
\end{equation}
Unlike in a stellar atmosphere, energy is locally produced so that the flux is
not constant.

Integrating the second moment equation Eq.~(\ref{eq:diffK}) over $m$, one
derives: 
\begin{align} 
   K              &= K(0) + H(0) \,\,\left( \tau_H -\Delta\tau_H \right) 
                     \label{eq:K}\\
   \mathrm{with}
   \quad\tauh (m) &= \int_{0}^{m} \chih(\zeta)               \diff{\zeta}\\
   \mathrm{and} 
   \quad\dtauh(m) &= \int_{0}^{m} \chih(\zeta) \theta(\zeta) \diff{\zeta} 
                     \label{eq:tauth}
\end{align}
where $\tauh$ and $\dtauh$ are respectively the flux weighted and the
``energy dissipation weighted'' mean optical depths.  In a stellar
atmosphere, one has $ K = K(0) + H(0) \tauh $.  Here, $\dtauh$
represents a correction to the flux-weighted mean optical depth due to
the decrease of the flux toward the mid-plane of the disk.  In the
large depth approximation, we have: ${\rm d\,}T^4/{\rm d}\tauh \propto
H$.  Since in a stellar atmosphere, the flux $H$ is constant, $T^4
\propto H \tauh$, whereas in a disk $H$ decreases with $\tauh$ until
$T^4 \leq H(0) \tauh$, hence the corrective term $\dtauh$.

We then introduce the Eddington factors,
\begin{equation}
\fK = \frac{K}{J} \quad {\rm and} \quad \fH = \frac{H(0)}{J(0)}.
\label{eq:Edd}
\end{equation}
In the large depth approximation, met in thick parts of the disk, they are 
respectively equal to 1/3 and 1/2.

\subsubsection{Formal solution}
\label{sec:model.solution}

Combining Eq.~(\ref{eq:Tfuncu}), Eq.~(\ref{eq:K}) and
Eq.~(\ref{eq:Edd}), we get the temperature distribution:
\begin{equation}
      T^4 = \frac{\kappaj \Teff^4}{4\kappab \fK}
                \Bigg[
                   \left( \tauh-\dtauh + \frac{\fK(0)}{\fH} \right)
                + \frac{\fK}{M \kappaj}\,\frac{u}{U} 
                 \Bigg]
   \label{eq:formsol1}
\end{equation}
with $\sigma_B \Teff^4 = U M$.

In the case of strict thermal equilibrium, the intensity-weighted
and flux-weighted mean opacities $\kappaj$ and $\kappah$ can be substituted
by the Planck and Rosseland mean opacities $\kappab$ and $\kappar$.

Equation~(\ref{eq:formsol1}) means that the vertical distribution of
the temperature in a disk is similar to its distribution in a stellar
atmosphere:
\begin{equation}
   T^4(\tau) = \frac{\kappaj(\tau)}{4\kappab \tau \fK(\tau)} \Teff^4
               \left[ \tau + \frac{\fK(0)}{\fH} \right],
\end{equation}
with two main differences:
\begin{itemize}
  \item[(i)] a local term, proportional to the energy dissipation $u$ is added.
  In a stellar atmosphere no energy dissipation occurs in the intermediate
  layers.  The factor proportional to the density of energy per unit mass $u$,
  in Eq.~(\ref{eq:formsol1}) is also inversely proportional to the optical
  depth of the disk, $\kappaj M$.  This factor is therefore dominant in zones
  where the disk is thin.  
  \item[(ii)] the optical depth $\tau$ becomes $\tauh-\dtauh$.  The term
  $\dtauh$ is positive but cannot exceed $\tauh$ depending where the energy
  dissipation occurs in the vertical structure.  The two extreme situations are
  when the dissipation function is a Dirac function located either in the
  central plane of the disk, or at the top of the photosphere.  In the first
  situation, $\dtauh=0$ everywhere, whereas in the second one $\dtauh=\tauh$.
  Above the layers where most dissipation occurs, the correction $\dtauh$ can
  be ignored and the disk behaves like a stellar atmosphere: there is no local
  energy source and all radiation comes from deeper layers.  In the central
  layers, $\theta(m) \approx 1$ and $\tauh(m)-\dtauh(m)$ is constant: most
  energy dissipation occurs in upper layers, and, in other terms there is
  almost no radiation coming from deeper layers.  Therefore the temperature is
  more or less constant.  In the particular case where the energy is dissipated
  only in the central plane $\tauh-\dtauh=\tau_H$ and the disk photosphere
  behaves like a stellar atmosphere.  \end{itemize}
\subsubsection{Several sources of energy dissipation}
\label{sec:model.diffsources}

In the case where there are several sources of energy dissipation releasing
the energy $u_k(m)$ in the layer $m$, the formal solution given in
  Eq.~(\ref{eq:formsol1}) can be split as the sum of the
  temperature distributions (see appendix \ref{app:split}):
\begin{equation}
  \label{eq:formsol2}
        T^4(m) = \sum_{k} \, T_k^4(m),\\
\end{equation}
with the temperature distribution of source $k$, 
\begin{align}
  \label{eq:formsolk}
  \begin{split}
    T_k^4(m) = \frac{ \kappaj (\Teff)_k^4}
                    {4\kappab \fK}
                \Bigg[
                &\left( \tauh-(\dtauh)_k + \frac{\fK(0)}{\fH} \right)\\
                &+ \frac{\fK}{M \kappaj}\,\frac{u_k}{U_k} 
                 \Bigg],\\
  \end{split}
\end{align}
where the effective temperature of the $k$-th source is defined by:
\begin{equation}
    \begin{split}
    \sigma_B (\Teff)_k^4 &= \int_0^M \! u_k(\zeta) \diff{\zeta}\\
                                  &= U_k M.
    \end{split}
    \label{eq:Tk}
\end{equation}

The quantity $(\Delta\tau_H)_k$ is defined by Eq.~(\ref{eq:tauth}) where the
vertical distribution function $\theta(m)$ is replaced by the corresponding
$k$-th energy source,
\begin{equation}
\theta_k(m) = \frac{1}{U_k M} \; \int_0^m u_k(\zeta) \diff{\zeta}.
   \label{eq:thetak}  
\end{equation}

We see that when dealing with several sources of energy dissipation, one can
treat them separately from the analytical point of view (see
Sect.~\ref{sec:model.external}).  However in the full modelization, the field
of radiation which defines $J(m)$, $H(m)$, $\fK(m)$ and $\fH$ must be computed
globaly.


\subsection{Influence of external radiative sources}
\label{sec:model.external}

\citet{hub90} already took into account the effect of the irradiation by
the central star \citep[see also][]{hub91}.  However, when addressing the
influence of the radiation from external sources, we can use
Eqs.~(\ref{eq:formsol2}) and (\ref{eq:formsolk}) taking into account as an
additional source of energy dissipation the reprocessing of the incoming
radiation. To compute the reprocessed energy $u_\mathrm{A}$ and its
influence on $(\dtauh)_\mathrm{A}$, we need to solve the radiative transfer
of the direct and scattered incoming radiation.

The flux coming from a central source is highly directional and cannot be
processed as isotropic radiation. For convenience, one may split the
radiation field into several components, and consider separate partial
transfer equations for the individual components. Some of the radiation
components may be considred as a source of additional heating.  Only the
component containing the thermal emission term, $\kappa_{\nu}B_{\nu}$ can
be treated with the formalism developed in the previous section. The mean
opacities $\kappaj$, $\kappab$, $\chih$, and the mean Eddington factors
$\fK$ and $\fH$ are then defined as appropriate averages over the moments
of this particular component of the radiation field.

Here, we process separately (i) the radiation from internal heating sources
or from external light absorbed and reprocessed in the disk, a radiation
which is mainly isotropic and thermalized, and (ii) the radiation coming
from external sources which is not absorbed.  In this latter case, we again
split the incoming radiation into two terms, one attenuated coming
directly from its external source in a given known direction, and one that
has been scattered once or several times and that should be mainly
isotropic.

We use the following notations:
\begin{enumerate}
   \item $\Inu$: radiation emitted by the disk itself. (All thermal flux 
   related quantities are written without superscript.)
   \item $\Inus$: radiation coming from external sources and scattered in the 
   disk
   \item $\Inuz$: attenuated radiation coming from external sources
   \item $\uv$: the energy dissipated locally (e.g.\ viscosity process)
   \item $\ua$: the energy coming from the incident photons absorbed in the 
   disk and reprocessed
\end{enumerate}

This approach is quite similar to Chandrasekhar's one \citep{cha60} and was
used by \citet{mb91}.  It consists of writing the equations of transfer for the
total intensity $\Inu$ and then deriving equations involving non-local terms
$\Inuz$ and $\Inus$.  Since the radiative transfer of $\Inu$ and non-local
terms are treated separately, the coupling between them can be seen either as a
loss of energy for the non-local radiation, or as a gain of energy for $\Inu$.
The stellar flux can now be seen as energy dissipation similar to that due
to viscosity.

\subsubsection{Energy reprocessed by the disk}

In the presence of an external source of radiation, the energy dissipation term
$\ua$ linked to the reprocessing of the radiation coming from external sources
and absorbed in the disk is directly linked to the specific intensity of the
incoming radiation by:
\begin{equation}
\begin{split} 
    \rho(m) \ua(m) = 
    4 \pi 
    \int_{-1}^{+1} \!\!\! \int_0^{+\infty} \!\!  
    \kapnu (&\Inuz(m, \mu)\\
            & + \Inus (m, \mu))  \diff\nu \diff\mu 
\end{split} 
\end{equation}

With $\Jz$ and $\Js$ the frequency-integrated zeroth order moments of
$\Inuz$ and $\Inus$,
\begin{align}
   \Jz   &= \int_0^{+\infty} \!\! \Jnuz \diff\nu
   \quad \mbox{ with } \quad
   \Jnuz &= \int_{-1}^{+1} \!   \Inuz(\mu) \diff\mu \label{eq:Jnuz}, \\
   \Js   &= \int_0^{+\infty} \!\! \Jnus \diff\nu 
   \quad \mbox{ with } \quad
   \Jnus &= \int_{-1}^{+1} \!   \Inus(\mu) \diff\mu \label{eq:Jnus},
\end{align}
and, the intensity weighted opacities,
\begin{align}
   \kappajz  &= \int_0^{+\infty} \!\! (\kapnu/\rho) \Jnuz
   \diff\nu /\Jz,\\ 
   \kappajs  &= \int_0^{+\infty} \!\! (\kapnu/\rho) \Jnus
   \diff\nu / \Js. 
\end{align}
$\ua$ can be expressed as:
\begin{equation}
   \ua =  4 \pi (\kappajz \Jz + \kappajs \Js ),
\label{eq:ufuncrad}
\end{equation}
The energy absorbed $\ua$ is therefore proportional to the mean incident flux
and to the absorption coefficient of the medium.

\subsubsection{Radiative transfer of the incoming radiation}

We study the transfer of the radiative intensities $\Iz$ and $\Is$ upon
which our knowledge of $\ua$ and $\dtauha$ depends. The equations
  of transfer for the incoming radiation within the frame of coherent
  isotropic scattering:
\begin{eqnarray}
  \label{eq:radtransf0}
  \mu \pder{\Inuz}{z}(z,\mu) &= &-\chinu\Inuz (z,\mu),\\
  \label{eq:radtransfs}
  \mu \pder{\Inus}{z}(z,\mu) &= &-\chinu\Inus (z,\mu) 
  \begin{array}[t]{l}
    + \signu\Jnus (z,\mu)\\
    + \signu\Jnuz (z,\mu),
  \end{array}
\end{eqnarray}
where $\signu\Jnuz (z,\mu)$ accounts for the scattering of the
incoming radiation and $\signu\Jnus (z,\mu)$ for multiple scattering.

We define the direction and frequency averaged opacities,
\begin{align}
  \chijz   &=  \int_0^{+\infty} \!\! (\chi_\nu/\rho)   \Jnuz
  \diff\nu /\Jz; \\ 
  \chihz   &=  \int_0^{+\infty} \!\! (\chi_\nu/\rho)   \Hnuz
  \diff\nu /\Hz; \\ 
  \sigmajz &=  \int_0^{+\infty} \!\! (\sigma_\nu/\rho) \Jnuz
  \diff\nu /\Jz ; \\ 
  \chihs   &=  \int_0^{+\infty} \!\! (\chi_\nu/\rho) \Hnus
  \diff\nu /\Hs; 
\end{align}
where $\Jnuz$, $\Jnus$, and, $\Hnuz$, $\Hnus$ are respectively the
zeroth and first moments of $\Inuz$ and $\Inus$.
Therefore the frequency-integrated zeroth, first and second moments of
$\Inuz$ and $\Inus$, namely $\Jz$, $\Hz$, $\Kz$ for direct incoming
radiation and $\Js$, $\Hs$, $\Ks$ for scattered incoming radiation, are
linked by the following equations:
\begin{align}
   \label{eq:transfer0}
   \der{H^0}{m} &= -\chijz\Jz 
      &\quad \mathrm{and} \quad 
   \der{K^0}{m} &= -\chihz\Hz \\
   \label{eq:transfers}
   \der{\Hs}{m} &= -\kappajs\Js +  \sigmajz\Jz 
      &\quad \mathrm{and} \quad
   \der{\Ks}{m} &= -\chihs\Hs .
\end{align}

Therefore, the reprocessed energy,
\begin{equation}
  \ua(m) = - 4 \pi \big(\der{\Hz}{m}+\der{\Hs}{m}\big),
\end{equation}
leads to 
\begin{align}
   \Ua M           &= 4 \pi \big( \Hz(0) + \Hs(0) \big)\\
   \thetaa(m)      &= 1 - \frac{\Hz(m) + \Hs(m)}{\Hz(0) + \Hs(0)}
\end{align}
assuming $\Hs(M)=0$ and $\Hz(M)=0$, \latin{i.e.} assuming that the equatorial
plane is a plane of symmetry for the source of external radiation. This is
truly the case for a central star or a uniform ambiant medium.

Using the second moment of the incoming and reprocessed radiative
equations, we derive $\dtauha$,
\begin{align}
   \begin{split}
      \dtauha = \tau_H + 
      \frac{
          \omegaz \big[ \Kz-\Kz(0) \big] + \omegas \big[ \Ks-\Ks(0) \big]
      }{
         \Hz(0)+ \Hs(0)
      }
   \end{split}
\end{align}
where we define respectively $\omegaz$ and $\omegas$, the mean
ratios of $\chih$ to $\chihz$ and $\chihs$:
\begin{align} 
   \int_0^m \chih(\zeta) \Hz(\zeta) \diff\zeta &= 
      \omegaz(m) \int_0^m \chihz  (\zeta)   \Hz(\zeta) \diff\zeta\\
   \int_0^m \chih(\zeta) \Hs(\zeta) \diff\zeta &= 
      \omegas(m) \int_0^m \chihs(\zeta) \Hs(\zeta) \diff\zeta .
\end{align}
If the radiation is absorbed in a geometrically thin layer, where the
temperature and density are more or less constant, then $\omegaz \approx
\chih/\chihz$.  It is the ratio of the absorption of \emph{reprocessed}
radiation (generally in the thermal infrared) to the absorption of
\emph{incoming} radiation (often in the visible, or UV, or X-ray domain).  In
the case of a hot source like stellar irradiation or a UV field, this ratio is
expected to be less than unity.

Finally the contribution of the incoming radiation source to the temperature
distribution is: 
\begin{eqnarray}
    \Ta^4(m) &= 
    &\frac{\kappaj}{4\kappab \fK} \; \Teffa^4 \nonumber \\
  \label{eq:formsolA}&&
  \bigg[
      \frac{
         \omegaz \big[ \Kz(0)-\Kz \big] + \omegas \big[ \Ks(0)-\Ks \big]
      }{
         \Hz(0)+ \Hs(0)
      }\\&&
      + \frac{\fK(0)}{\fH} + \frac{\fK}{M \kappaj}\,
        \frac{\kappajz\Jz + \kappajs\Js}{\Hz(0) + \Hs(0)}
   \bigg] \nonumber
\end{eqnarray}

We show in Appendix \ref{app:incrad} how to compute $\Jz(m)$, $\Js(m)$,
$\Hz(m)$, and $\Hs(m)$ in the case of an external point-like source
where the irradiation from the central source is highly collimated.
As demonstrated in Sect.~\ref{sec:model.diffsources}, the temperature
vertical structure of a sum of point-like sources is the sum of the
temperatures computed for each source to the fourth power.

\subsubsection{Interpretation}
\label{sec:centinterp} 

In this section, we discuss in detail the physical interpretation of 
the complex Eq.\ (\ref{eq:formsolA}). Since we know how to add the
contributions of various energy sources by using Eqs.~(\ref{eq:formsol2})
and (\ref{eq:formsolk}), we can restrict ourselves to a point-like source
located at infinity without losing generality. An extended source 
will then  be processed as the sum of point-like sources.

The dependence of the central temperature to the incoming stellar radiation
is relatively simple. Since the topmost layer emits half upward to the
outside and half downward toward the inner layers, the latter ones receive
half of the emission.  Their effective temperature is then $\Ta / 2^{1/4}$.
Since there is no flux inside (for $\tau \gg 1$), the temperature is
constant and equal to this temperature. This intuitive result is compatible
with Eq.\ (\ref{eq:formsolA}). As a matter of fact, if $\muz$ is the cosine
angle of the incoming radiation and since $K = \muz H$, the first term in
the square brackets is of the order of $\omegaz\muz$. For a slanted
incident stellar radiation, we have $\muz \ll 1$ and therefore $\omegaz\muz
\ll 1$.  Deep inside the disk, the third term, evanescent since the
incoming radiation is absorbed, is also negligible. The brackets reduce to
$\fK(0)/\fH \approx 2/3$.  The central temperature is therefore smaller
than the effective temperature $\Ta$ by a factor $\approx 2^{-1/4}$. In
conclusion, half of the absorbed energy is irradiated by the hot topmost
layer and another half by the cooler inner layers.

At the surface, the third term in the square brackets reduces to $1/(\muz
\omegaz)$ by using $\Jz = H/\muz $ and is dominant.  The temperature is
therefore much higher than the effective temperature. This fact is explained in
\citet{mb91}: the topmost layer, vertically optically thin but radially
optically thick, absorbs almost all stellar light. This layer is superheated by
a factor of $(1/\muz)^{1/4}$. \citet{cg97} refined this view by multiplying the
term $1/\muz$ by $1/\omegaz$.  The first factor is a pure geometrical effect of
the slanted incidence whereas the second one comes from the difference between
the opacities for the visible incoming radiation and the infrared reprocessed
radiation.

\subsubsection{Overview of other heating sources}
\label{sec:htr}

The flux radiation from the ISM can be processed on the same basis as the
stellar irradiation.  The main difference is that the ISM does not
irradiate in a particular direction.  If we assume that it consists of an
isotropic black body radiation of characteristic temperature $T_{\rm ISM}$,
we can apply Eq.~(\ref{eq:TRmb}) within the two-stream approximation
(\latin{i.e.} with $ \mu^* = 1/\sqrt{3} $ using the definition given in
Sect.~\ref{sec:ttscentral}) and we can demonstrate that the contribution of
the ISM radiation is a constant.  In fact, if the ISM is the unique source
of irradiation, the disk would be isothermal.

All other radiative sources can be treated exactly as the central star.  If
they are extended, the previous results derived for point-like sources are
integrated over angles.

The radial extent of the disk is much larger than the vertical one.  We can
therefore apply the approximations of large depth to the horizontal transfer.  This hypothesis,
combined with the LTE approximation, allows us to write the horizontal flux
$H_r$ as:
\begin{equation}
   H_r = - \frac{4}{3 \pi (\rho \kappar)} T^3 \pder {T}{r}
\end{equation}
The local energy gain for the material between $r$ and $r+{\rm d}r$ is $4
\pi (H_r(r+{\rm d}r) - H_r(r))$. One then derives the heating per unit of
mass of material that allow us to self-consistently compute the correction to
the vertical transfer, in the case $H_r \ll H$:
\begin{equation}
   u_H = \frac{4\pi}{\rho} \pder{H_r}{r} \quad
\end{equation}


\section{Application to T Tauri disks}
\label{sec:ttsdisks}

 The two main heating sources in the case of T Tauri disks are the
 viscous dissipation and the stellar light absorption.  We apply our
 general equations in this frame.

\subsection{Approximations}
\label{sec:radeq.approx}

All the terms $\tauh$, $\kappah$, etc.,  are dependent on the structure.
Moreover $\fK$ and $\fH$, are not known a priori.  Therefore we will seek
a self-con\-sis\-tent iterative method able to determine these quantities.

We now make two approximations concerning the opacities: the diffuse
intensity is a Planck distribution at the local temperature and the
radiation coming from a external source is also a Planck distribution at
the temperature of the source $T^*$.  We then derive:
\begin{itemize}
   \item $\kappaj  = \kappab$, the Planck mean opacity
   \item $\kappajz = \kappajs = \kappab^*$, the mean opacity for a black body 
   radiation at $T^*$ through a medium at temperature $T$.
   \item $\kappah  = \kappar$, the Rosseland mean opacity
\end{itemize}
Since the scope of this paper is to present analytical solutions and simple
numerical applications, we use the gray Rosseland opacities given by
\citet{bl94}. We also use the gray approximation $\kappab = \kappab^* =
\kappar$ and ignore diffusion.  Therefore $\omegas = \omegaz$ that we now
call $\omega$

The  $\fH$ and $\fK$ values will be close to 1/2 and 1/3.  These approximations
are only valid at large depths, but we refine these values in optically thin
zones at each iteration, as well as for $\omega$.

\subsection{Vertical temperature distribution}

\subsubsection{Viscous heating at each layer in the disk}

We consider here an energy density per mass unit which depends on local
conditions.  The energy dissipation per mass unit due to viscosity is given
by \citet{fkr85}:
\begin{equation}
   \uv(m) = \nuvisc(m) \left( r \der{\Omega}{r} \right)^2
\end{equation}
In the case of a thin Keplerian disk we get:
\begin{equation}
   \uv(m) = \frac{9}{4} \Omega^2 \nuvisc(m)
\end{equation}

Since $\uv(m)$ depends vertically only on $\nuvisc(m)$, the $\thetav$ function
depends only on the vertical structure of the viscosity \nuvisc{} and of its
mass-average $\bar{\nu}_{\rm visc}$.

Up to now there has been no exact description of the viscosity induced by the
turbulence.  In this paper, we have considered two different types of
viscosity: 

\begin{enumerate}\setlength{\itemsep}{1em}
   \item The Shakura-Sunyaev prescription stating that the viscosity is
   proportional to the local sound speed and local height scale:
   \begin{equation}
      \nu_{\rm  visc} = \alpha c_s  h \propto T
   \end{equation}
   The $\thetav$ function is therefore the distribution function of the
   temperature. The term $\tauh-\dtauhv$, by definition smaller than $\tauh$,
   can be much smaller if the temperature peaks at the surface of the
   photosphere.
 \item The $\beta$-prescription, derived from laboratory experiments, and
   applied to accretion disks by \citet{hure00}, states that
   \begin{equation}
       \nu_{\rm visc} = \beta \Omega r^2
   \end{equation}
   where $\beta$ is a constant ranging from $10^{-5}$ to $10^{-3}$. Here the
   viscosity is uniform along the vertical axis and does not depend on the
   temperature distribution of the disk.
\end{enumerate}

\subsubsection{Heating by the central stellar source}
\label{sec:ttscentral}

With the approximations of Sect.~\ref{sec:radeq.approx}, there is no reason to
distinguish the two specific intensities $I^0$ and $I^s$. If we call $I^* =
I^0 + I^s$, then we find an analytical solution very similar to Eq.~(28)
of from \citet{mb91}:
\begin{eqnarray}
   T_\mathrm{A}^4 &= \displaystyle 
   \frac{\pi \kappaj H^*(0)}{\fK \kappab \sigma_B}
   \bigg[ 
      &\omega^* \frac{K^*(0)-K^*}{H^*(0)} \nonumber\\
         &&+ \frac{\fK(0)}{\fH}
         + \frac{\fK \kappa^*_J J^*}{\kappa_J H^*(0)}
      \bigg]
\end{eqnarray}
where the quantities linked to the star incoming radiation are
superscripted with $*$. The only difference is the presence of $\omega^*$
which is not a priori equal to 1 as noted by \citet{dal98}.

We derive $J^*(0)$ and $H^*(0)$ from the geometrical properties of the system
using the approximation described by \citet{rp91}.  The star is seen by the
disk as a point-like source. It is valid beyond a few stellar radii; in the
inner parts this is not the case, nevertheless reprocessing is seldom dominant
in this region \citep[see Fig.\ 3 of][]{dal98}. A point-like source equivalent
to the star must present the same values for $J^*(0)$ and $H^*(0)$.  The
incidence angle is given by:
\begin{equation}
   \mu^* = H^*(0)/J^*(0)
\end{equation}

One finally obtains, using the gray optical depth $\tau$, the same expression
(33) of \citet{mb91} except for the factor $\omega^*$ already discussed in
the previous section,
\begin{eqnarray}
   T_\mathrm{A}^4 &= \displaystyle \frac{\Teffa^4}{4 \fK}
      \bigg[ 
           &\mu^*\omega^* \; \left(1- {\rm e}^{-\tau/\mu^*} \right) \nonumber\\
         &&+ \frac{\fK(0)}{\fH}
         + \fK \frac{{\rm e}^{-\tau/\mu^*}}{\mu^*}
      \bigg]
                     \label{eq:TRmb}
\end{eqnarray}
where the effective temperature of the stellar reprocessing is:
\begin{equation}
   \Teffa^4 = 4 \pi H^*(0) / \sigma_B
\end{equation}

\subsection{Equation of state --- hydrostatic equilibrium}

The equation of state in the disk is:
\begin{align}
                     P       &= f_{\rm gas} c_{\rm s}^2 \rho \label{eq:eqst}\\
   {\rm with}\quad c_{\rm s} &= \sqrt{\frac{k_{\rm B}T}{{\cal M}m_{\rm H}}}
\end{align}
First of all, radiation pressure is ignored in Eq.~(\ref{eq:eqst}) as suggested
by \citet{ss73}.  If the disk is assumed to be homogeneous and mixed by the
turbulence, the mass fraction of gas, $f_{\rm gas}$, is a known function of $T$
and $\rho$, and might even be a constant if evaporation is ignored.  The dust
is only a few percent of the mass of the disk, so we use $f_{\rm gas} = 1$ in
the present work.

We also assume that the disk dynamics is dominated by the central star, so that
it is Keplerian.  Then, in the thin approximation, $\Gz$ is a linear function
of $z$.  Combined with Eq.~(\ref{eq:eqst}), the hydrostatic equilibrium and the
local height scale become:
\begin{align}
   \der[2]{P}{m}           &= - \frac{c_{\rm s}^2 \Omega^2}{P}\\
   h                       &=   \frac{c_{\rm s}}{\Omega}\\
   {\rm with} \quad \Omega &=   \sqrt{\frac{GM*}{r^3}}
\end{align}


\subsection{Self-similar accretion model}
\label{sec:results.model}

The radial distribution of energy dissipation for the accretion $\Uv(r)$ and
the surface density $\Sigma(r)$ are essential parameters for the initialisation
step of our code. In order to be as general as possible, we use a prescription
with a self-similar accretion rate, instead of the model by \citet{ss73} which
assumes a uniform accretion rate over the disk.  We assume that some ejection
process removes material away from the close environment of the star within a
radius $r_0$.  A more detailed description of this mechanism can be found in
\citet{fer95}. The accretion rate is parametrized as:

\begin{equation} 
   \mdot(r) = \begin{cases} 
      \displaystyle \mdot_{\infty} \left( \frac{r}{r_0} \right)^\xi 
         &\mbox{if}\quad r < r_0\\[5pt] 
      \displaystyle \mdot_{\infty} 
         &\mbox{if}\quad r \ge r_0 
   \end{cases} 
\end{equation}
where $0\leq\xi\leq1$ is the ejection index \cite[see][]{fer95} and $r_0$ the
cut-off radius.  When $\xi=0$, this model corresponds exactly to the standard
accretion model by \citet{ss73}.

The dissipation per area unit $U_\mathrm{V}(r)$ and the mass column $\Sigma(r)$
can then be expressed as:
\begin{align}
   U_\mathrm{V}(r)    &= \frac{ 3 {\cal G} M_* \mdot(r) } { 8 \pi r^3}  f(r)\\
   \Sigma(r)          &= \frac{\mdot(r)}{3 \pi \nuvisc}                 f(r)
   \label{eq:sigma}
\end{align}

where $f(r)$ is a factor close to unity at large $r$.  Using the same method as
\citet{ss73} we obtain:
\begin{equation}
f(r) =
   \begin{cases}
      \displaystyle \frac{1}{1+2\xi} \left[
         1 - \left(
            \frac{r_*}{r}
         \right)^{\xi+\frac{1}{2}}
      \right] 
                        &\mbox{if $r < r_0$}\\[10pt]
      \displaystyle
      1 - \frac{2\xi}{1+2\xi} \sqrt{\frac{r_0}{r}} 
        - \frac{   1}{1+2\xi} \left(\frac{r_*}{r_0}\right)^\xi 
      \hspace{-5pt} \sqrt{\frac{r_*}{r}}
                        &\mbox{if $r \ge r_0$}
   \end{cases}
\end{equation}
%

\subsection{The numerical code}

Our set of equations are coupled and non-linear. They need to be solved 
numerically, within the framefork presented below. 

{\itshape The grid:} we use a logarithmic radial grid.  For each radius we
chose a non-uniform vertical grid so that the mass column coordinate $m(z)$ is
logarithmically sampled.

{\itshape Initial conditions:} from the radial structure described
in Sect.~\ref{sec:model.init}, we compute an isothermal vertical structure. 

{\itshape Iterative method:} Knowing the temperature and mass column, the
hydrostatic equilibrium is solved for. We deduce the density and pressure
distributions.  Then the temperature and optical depth are computed again using
the development of Sect.~\ref{sec:model.radiat}.  Subsequent quantities are
derived, like the kinematic viscosity. The value of $M(r)$ is refined so that
it matches Eq.~(\ref{eq:sigma}) using the average kinematic viscosity.  The
mass column grid is computed again with the new value of $M(r)$.  The latter
step is performed as many times as necessary to reach convergence (typically in ten
iterations).

The parameters used for the computations are reported in Table~\ref{tab}.
\def\eq{$=$}
\begin{table}[t]
  \centering
  \caption{Parameters used in the model}
  \label{tab}
   \begin{tabular}{|rcl|rcl|}
     \hline
      $r_*$         & \eq &$ 2\,R_\odot   $ &$M_*        $ &\eq &$ 0.5\,M_\odot$\\
      $T_*$         & \eq &$ 4000\,{\rm K}$ &$T_{\rm ISM}$ &\eq &$ 15\,{\rm K} $\\
      $r_0$         & \eq &$ 100\,r_*     $ &$\alpha     $ &\eq &$ 0.01        $\\
      $r_{\rm min}$ & \eq &$ 2.2\,R_\odot $ &$A_{\rm V}  $ &\eq &$ 1           $\\
      $d$           & \eq &$450\,\mbox{pc}$ &              &    &               \\
     \hline
   \end{tabular}
\end{table}


\section{Results and discussion}
\label{sec:results}

In this paper, we chose to treat mainly the radiative transfer in the disk.
Even if we have developed the formalism of the central source heating, we will
ignore its contribution in a first step and concentrate only on the
viscous dissipation.  In a future paper, we will present the case of
heating by the central star.  Moreover the heating by the central star becomes
dominant at radii larger than a few astronomical units as shown in Fig.\ 3 of
\citet{dal98} so the results presented here will be correct only for the inner
part of the disk and therefore for the visible and infrared thermal domain.

First, we compute the physical conditions in the disk: temperature, density,
optical depth.  Then we use these physical conditions to compute astronomical
observables like SEDs, millimetric images and interferometric complex
visibilities and direct images.

\subsection{Physical conditions in the disk}

\subsubsection{Temperature and density}

\begin{figure*}[t]
  \parbox[t]{0.48\hsize}{
    \includegraphics[width=0.95\hsize]{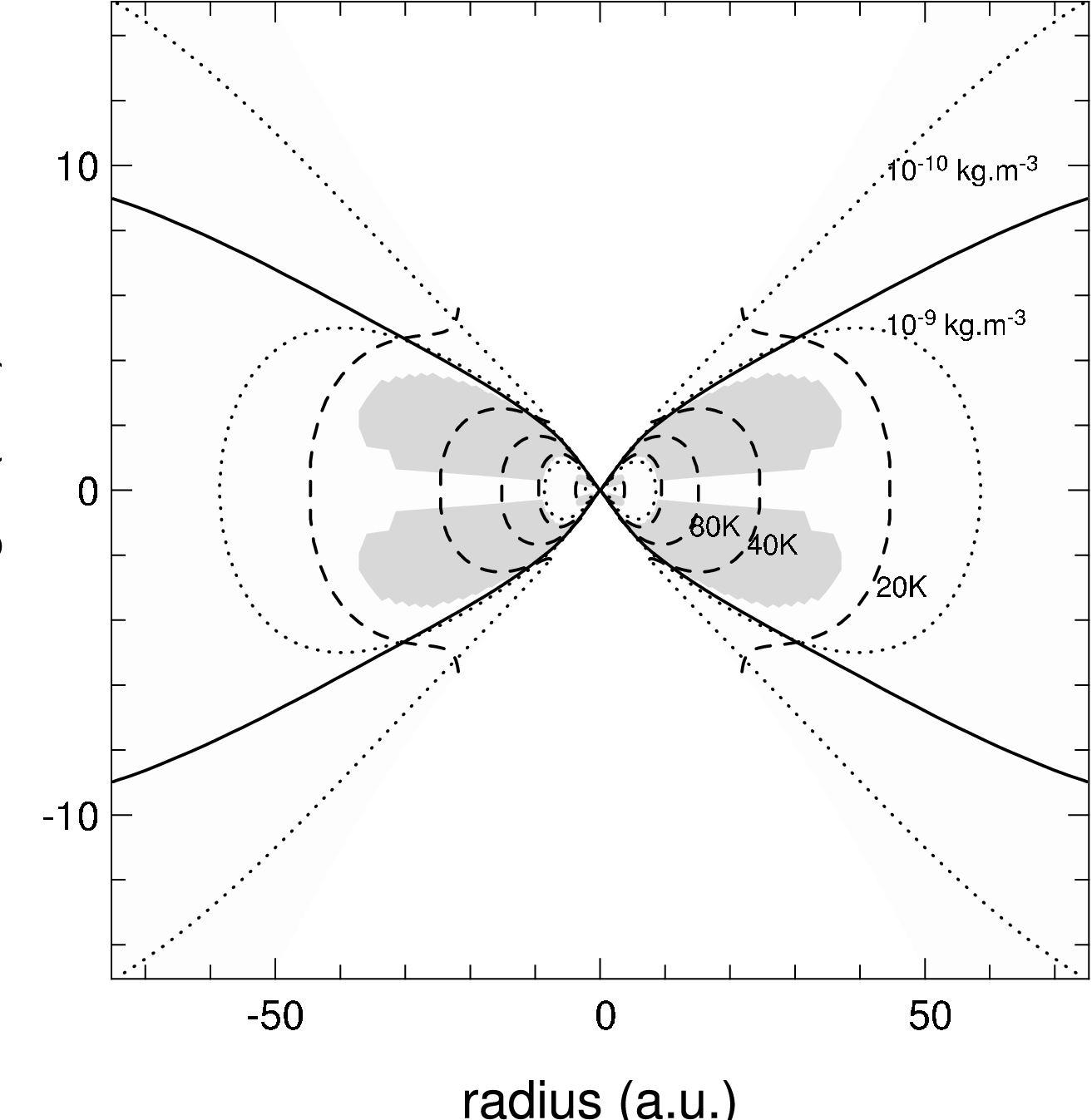}
    \caption{Iso-contour maps of a T Tauri disk with $\mdot = 10^{-5}\,\msyr$.
      \textit{Dashed lines:} lines of equal temperature; 
      \textit{Dotted lines:} lines of equal density; 
      \textit{Solid line:} limits of the optically thick region ($\tau = 1$);
      \textit{Gray zone:} zone of instability with respect to the convection.}
    \label{fig:map}
    }
  \hfill
  \parbox[t]{0.48\hsize}{
    \includegraphics[width=0.95\hsize]{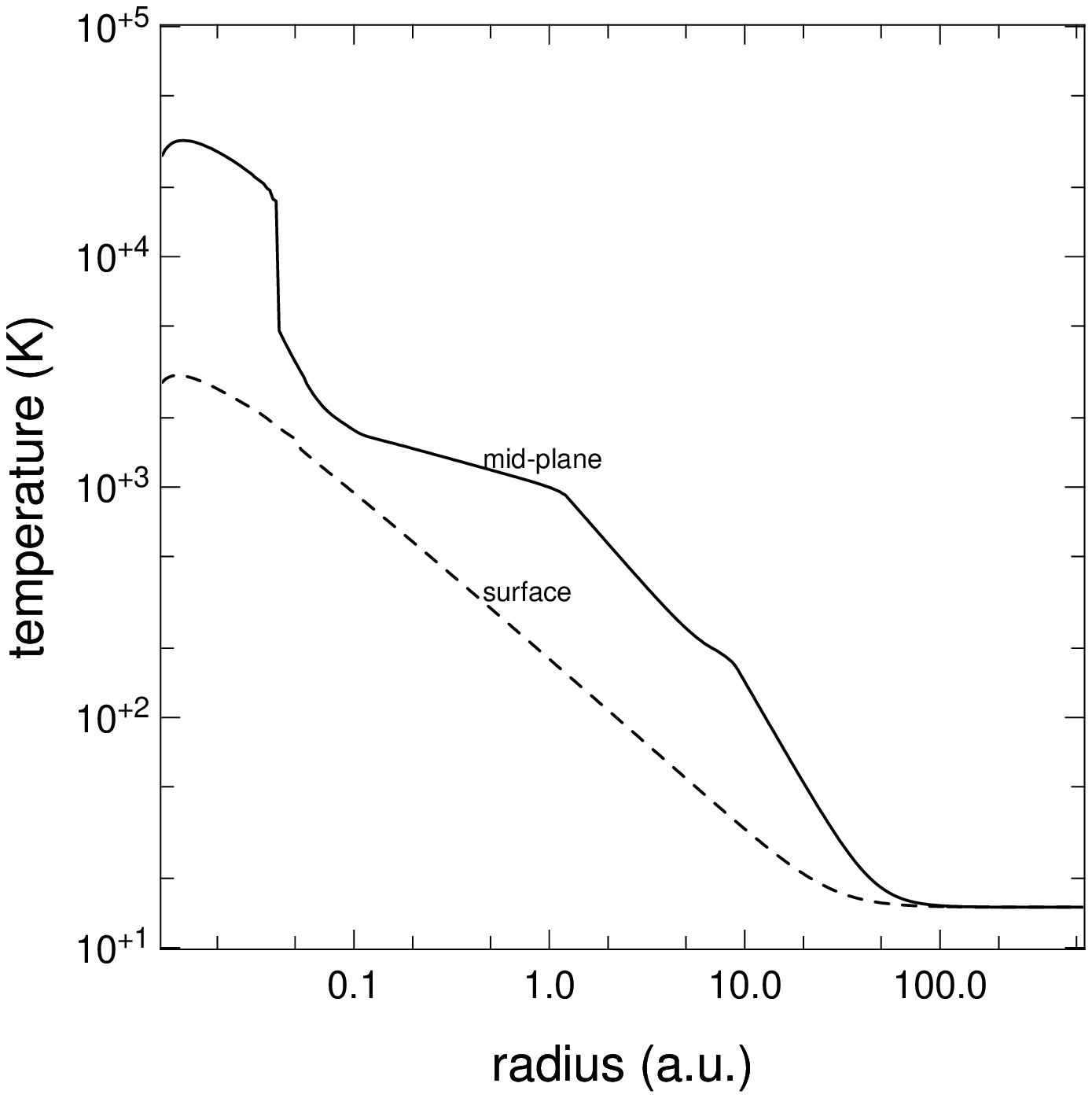}
    \caption{Radial distribution of the temperature in the central layer
  and of the effective temperature (located at $\tau=1$).}
    \label{fig:temp}
    }
\end{figure*}

A map of density and temperature for a standard disk ($\xi = 0$) with an
accretion rate of $10^{-5}\,\msyr$ is displayed in Fig.~\ref{fig:map}.  The
shape of the temperature iso-contours illustrates that the disk temperature is
almost constant in the optically thick part, whereas the temperature changes
quickly with the altitude in the upper layer.  As previously guessed, the disk
appears roughly isothermal over the vertical dimension in its inner part. 

Figure \ref{fig:temp} displays the variation of the temperature with the
radius. We verify that the effective temperature, \latin{i.e.} at $\tau=1$, is
proportional to the dissipated flux: it follows an $r^{-3/4}$ law.  We stress
that the central temperature is much higher than the effective temperature.

We check that the disk is optically thick even at large radii while the
geometrical thickness of the disk remains much smaller than unity: if we define
the disk surface as the region of optical depth unity, we find $z/r \approx
0.1\mbox{--}0.2$.  

The geometry of the density contours displayed in Fig.~\ref{fig:map} have a
different shape depending whether they are located in the central optically
thick part or in the upper optically thin layers. Figure~\ref{fig:map} also
displays the disk parts which are unstable to convection by computing
the Schwarzschild criterion. For the time being, the consequences of this
convection are not taken into account in our model.

\subsubsection{Column density}

\begin{figure*}[t]
   \parbox[t]{0.48\hsize}{
       \includegraphics[width=0.95\hsize]{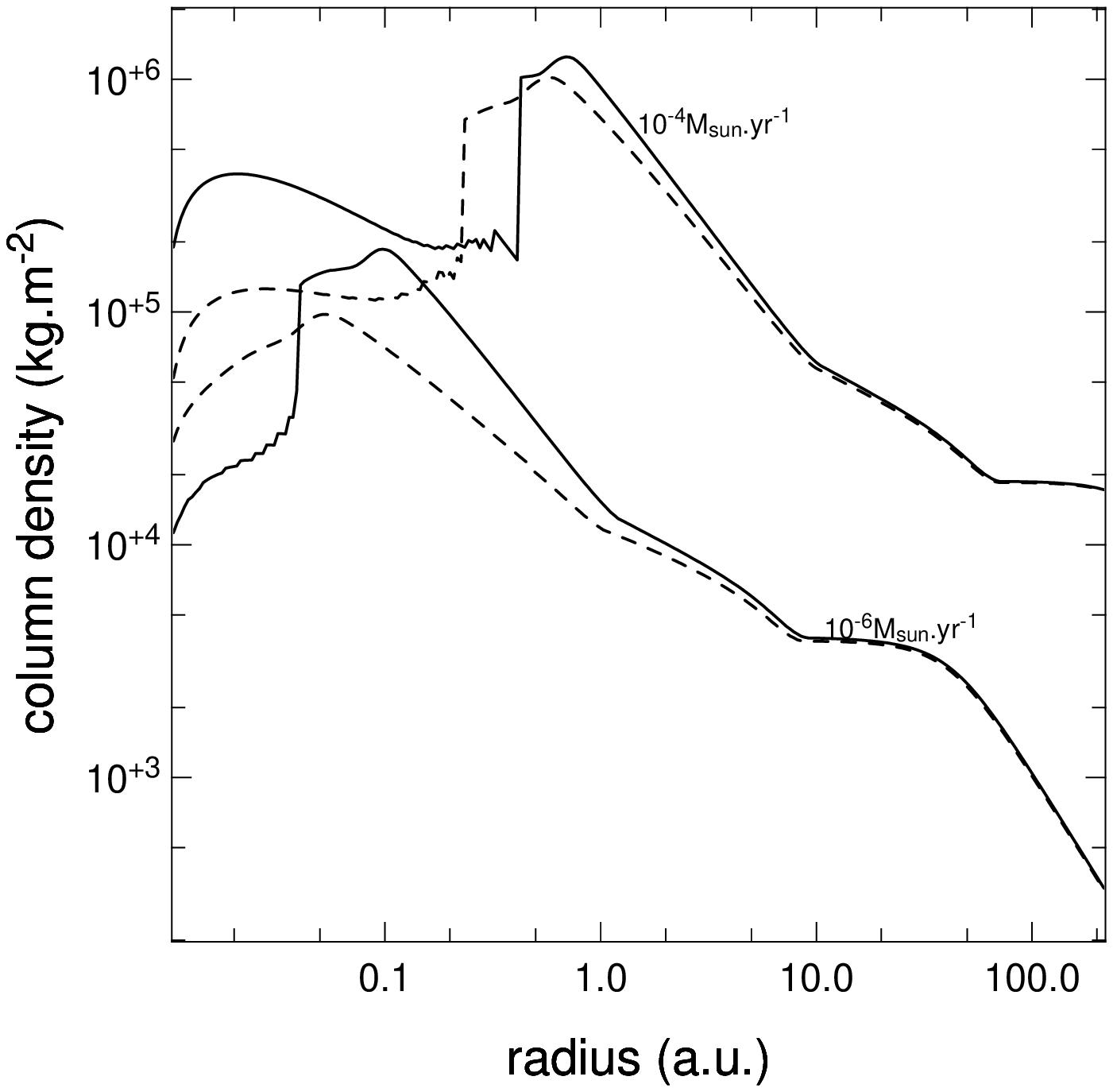}
       \caption{Column density in an $\alpha$-disk for two different accretion rates.
          \textit{Solid lines:} $\xi = 0$; 
          \textit{Dashed lines:} $\xi = 0.5$}
       \label{fig:sigma}
   }
   \hfill
   \parbox[t]{0.48\hsize}{
      \includegraphics[width=0.95\hsize]{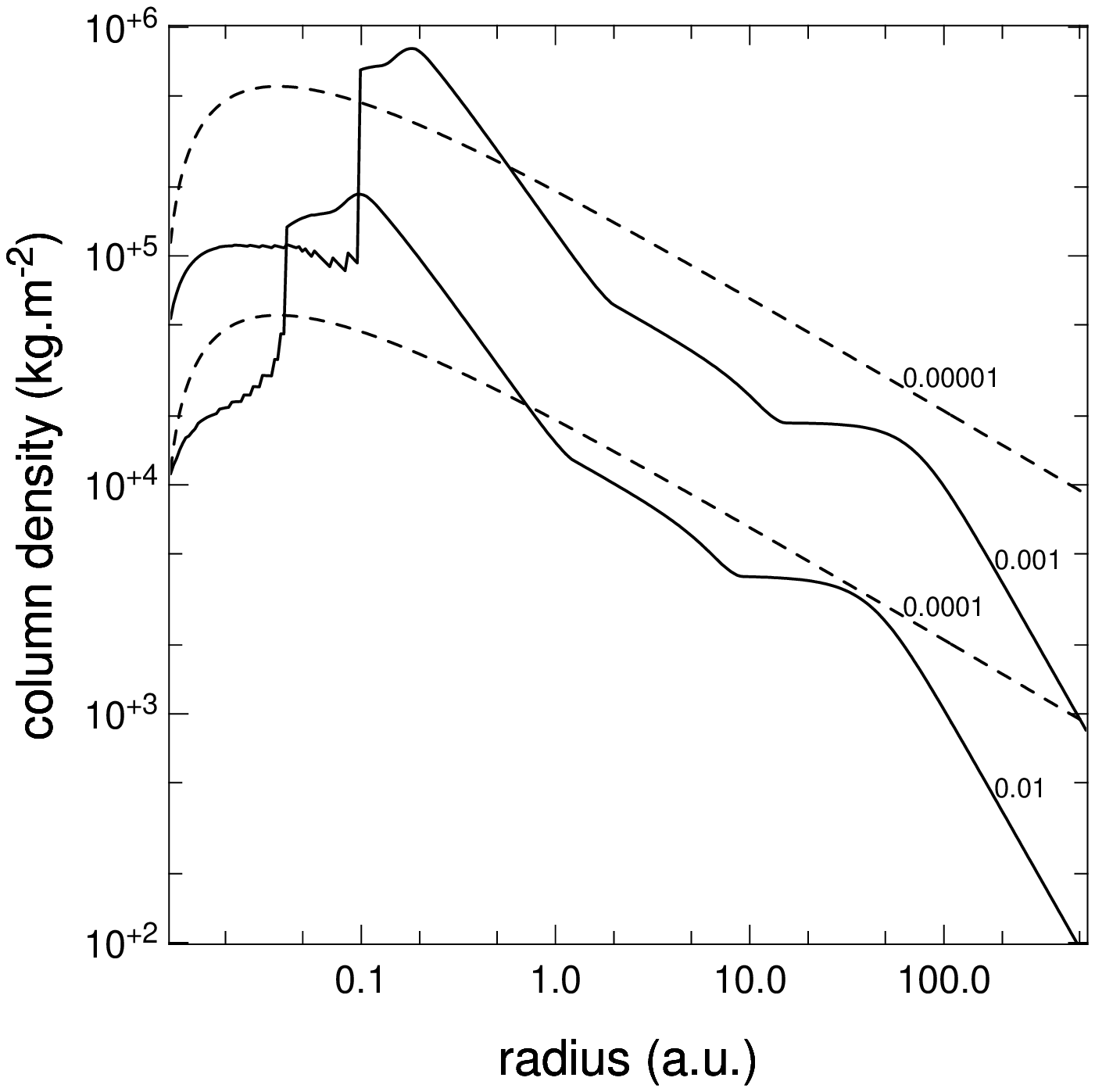}
      \caption{Column density in a disk with different viscosity prescriptions.
         \textit{Solid lines:} $\alpha = 10^{-3},\:10^{-2}$;
         \textit{Dotted lines:}  $\beta  = 10^{-4},\:10^{-3}$} 
      \label{fig:alphabeta}
   }
\end{figure*}

Figure~\ref{fig:sigma} shows the mass column density as a function of the
radius in the case of a standard $\alpha$-disk with two values of the ejection
parameter $\xi$.  We take the cut-off $r_0$ at 100 stellar radii, \latin{i.e.}
about 1 AU. Two values of the accretion rate have been considered: $10^{-6}$
and $10^{-4}$\,$M_{\odot}$/yr. As expected, the radial structure of $\Sigma(r)$
changes in the close environment of the star when $\xi\neq0$.

Power-laws are often used by observers in the description of the radial
structure of disks, though one can see in Fig.~\ref{fig:alphabeta} that the
column density of an $\alpha$-disk cannnot be described with such a law.
However, on restricted domains of radius, the mass column in an $\alpha$-disk
can be described by the equation
\def\md{\ensuremath{{\dot{m}_6}}}
\begin{equation}
   \Sigma(r) \approx \Sigma_i \, (r/1\,\mbox{AU})^{-q_i}\,\mbox{$\mbox{kg.m}^{-2}$}
   \mbox{ for } r_i \le r \le r_{i+1}
   \label{eq:powlaw}
\end{equation}
where the coefficient $q_i$ ranges from -0.4 to 1.5 and $\Sigma_i$ is a 
constant.  These coefficients are given in Table~\ref{tab:ab} for two values
of the accretion rate.
\begin{table}[t]
   \caption{Parameters computed from the model to approximate the surface
     density by a power law by pieces, sorted by decreasing order of radius 
     domain.} 
   \begin{center}
   \begin{tabular}{|rrr|rrr|}
     \hline
   \multicolumn{3}{|c|}{$\mdot = 10^{-6}\,\msyr$} 
                                  & \multicolumn{3}{c|}{$\mdot = 10^{-4}\,\msyr$}\\
   $\Sigma_i$  &  $q_i$ &   $r_i$ & $\Sigma_i$   &  $q_i$ & $r_i$\\
   \hline
   $1  \,10^6$ & $ 1.5$ &    40    &         --- &    --- &  --- \\
   $4  \,10^3$ & $ 0.0$ &     9    & $2  \,10^4$ & $ 0.0$ &   67 \\
   $1.5\,10^4$ & $ 0.6$ &     1    & $2.5\,10^5$ & $ 0.6$ &    5 \\
   $1.5\,10^4$ & $ 1.2$ &     0.11 & $9  \,10^5$ & $ 1.4$ & 0.75 \\
   $4  \,10^5$ & $-0.3$ &     0.06 & $1.5\,10^6$ & $-0.5$ & 0.45\\
     \hline
   \end{tabular}
   \end{center}
   \label{tab:ab}
\end{table}

We show in Fig.~\ref{fig:alphabeta} the influence of $\alpha$ on the structure
of the disk. These results are compared with the ones obtained with the $\beta$
viscosity prescription. The general behaviour is somewhat different except for
the $1-10\mbox{ AU}$ region. Therefore observing inside 1 AU or outside 10 AU
could be decisive in determining the viscosity law.

\subsection{Astronomical observables}
\label{sec:observables}

One of the aims of this work is to compute physical conditions around a young
star so that we can calculate the electromagnetic field emerging from this
region.  This emission can then be analysed by several types of instruments:
photometers, spectrographs, imaging cameras, interferometers, etc. The next
sections give examples of astronomical observables that can be used to
constrain our knowledge of the circumstellar environment.

\subsubsection{Spectral energy distribution} 

\begin{figure*}[t]
     \parbox{0.48\hsize}{
      \includegraphics[width=0.95\hsize]{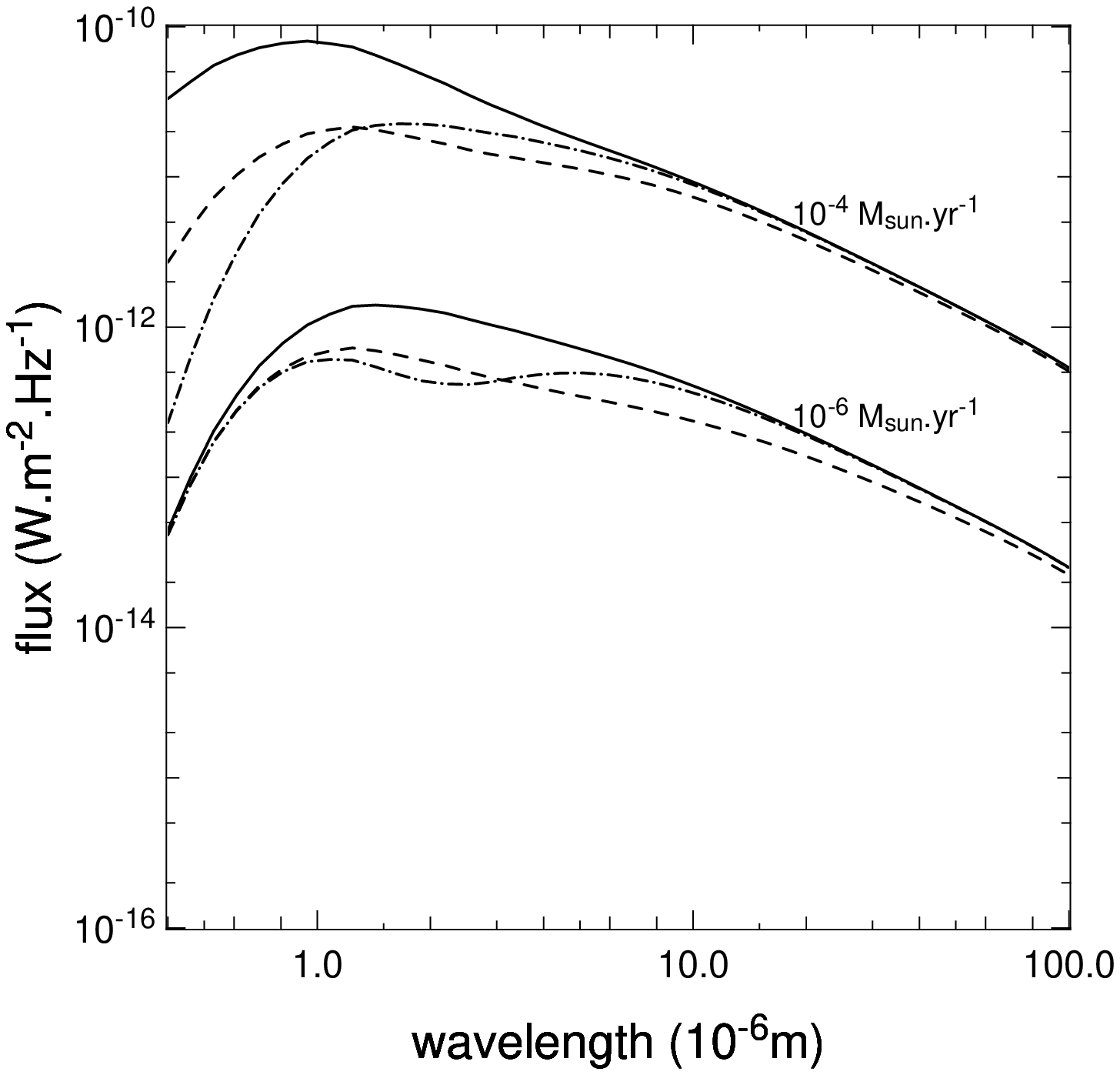}
      \caption{Spectral energy distribution ($\lambda F_\lambda$
      vs. $\lambda$) for different disk models
         \textit{Solid lines:} standard disk;  
         \textit{Dashed lines:} disk with $\xi = 0.5$; 
         \textit{Dashed \& doted lines:} standard disk with an inner hole
         of $8 r_*$.} 
      \label{fig:sed1}
   }
   \hfill
   \parbox{0.48\hsize}{
      \includegraphics[width=0.95\hsize]{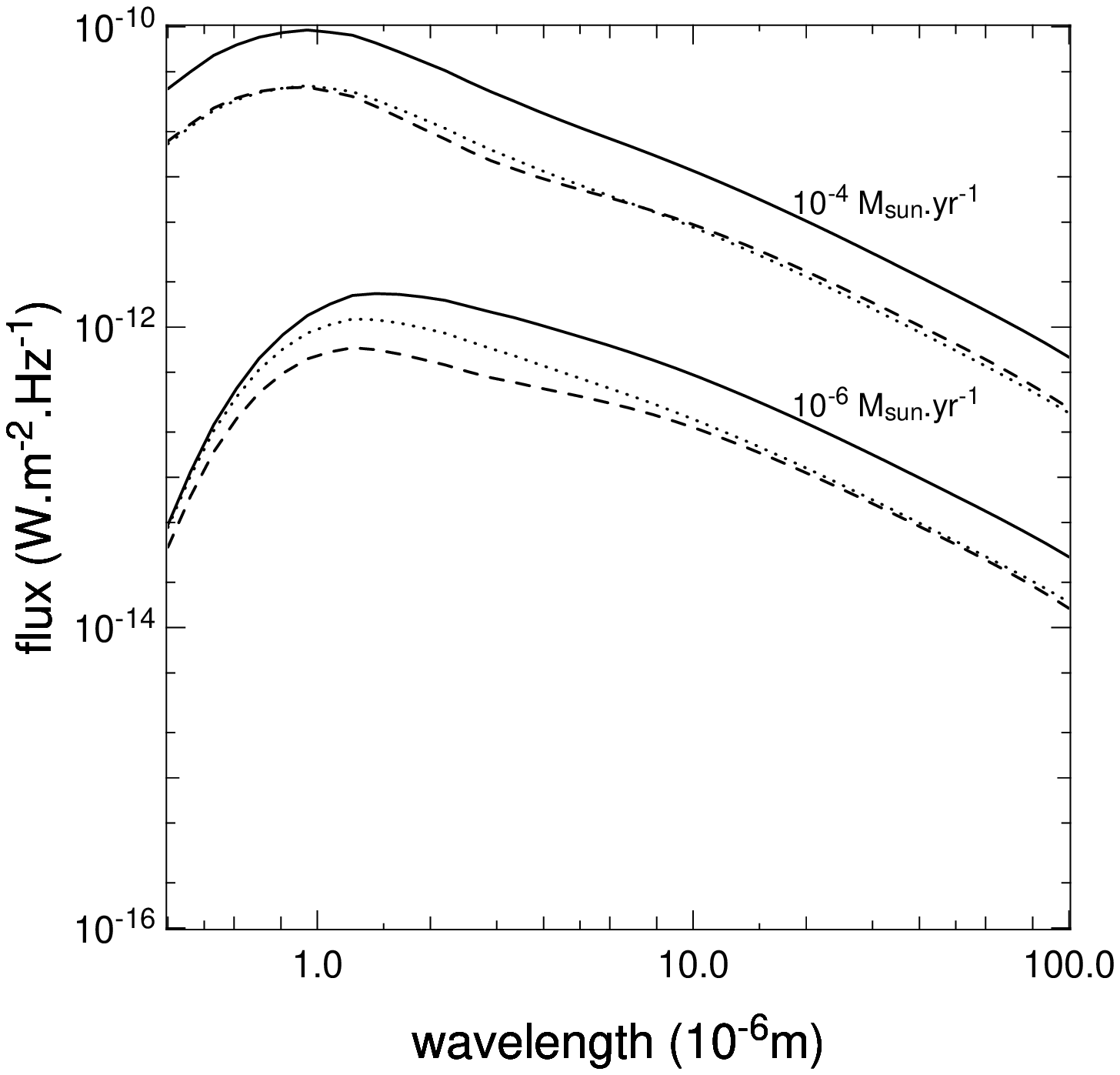}
      \caption{Spectral energy distribution ($\lambda F_\lambda$
        vs. $\lambda$) of the disk model as a function of inclination.
         \textit{Solid lines:} pole-on disk.
         \textit{Dashed lines:} inclination of $i=45^o$. 
         \textit{Dotted liness:} same inclination without the flaring
         effect (see text for details).}
      \label{fig:sed2}
   }
\end{figure*}

The spectrophotometry technique gives access to the spectral energy
distribution.  We integrate along the line of sight the contribution of each
disk layer to the emergent flux. In the examples taken for Fig.~\ref{fig:sed1},
the SED is significantly altered in the range $1$--$5$\,$\mu$m with the
self-similar accretion model. The inner hot regions being depleted, the short
wavelengths are less present.  We assumed an overall extinction of $A_{\rm V} =
1$ in the visible and an inner radius of the disk of $r_{\rm min} = 1.1$
stellar radius. However, Fig.\ \ref{fig:sed1} shows that the SED is
significatively different from the one expected from a standard model with a
large inner gap.

Figure~\ref{fig:sed2} shows the influence of disk flaring on the
intermediate wavelength range of the SED. The difference between the two
curves representing the SED of a $45^o$-inclined disk lies in the location of
the photosphere, \latin{i.e.} the location of the disk layer that emits most of
the emergent flux. In the flaring case, the flux comes from the layer where
$\tau=1$, whereas in the so-called no flaring case, the flux comes from the
disk equatorial plane but with the same temperature as the effective
temperature. This latter case corresponds to the standard radial model where the
vertical structure is not taken into account. Even without the reprocessing of
the stellar photons, the effect of the vertical extension of the disk
atmosphere is therefore not negligible at intermediate wavelengths.

\subsubsection{Millimetric images}

\begin{figure}[t]
  \centering
  \includegraphics[width=0.95\hsize]{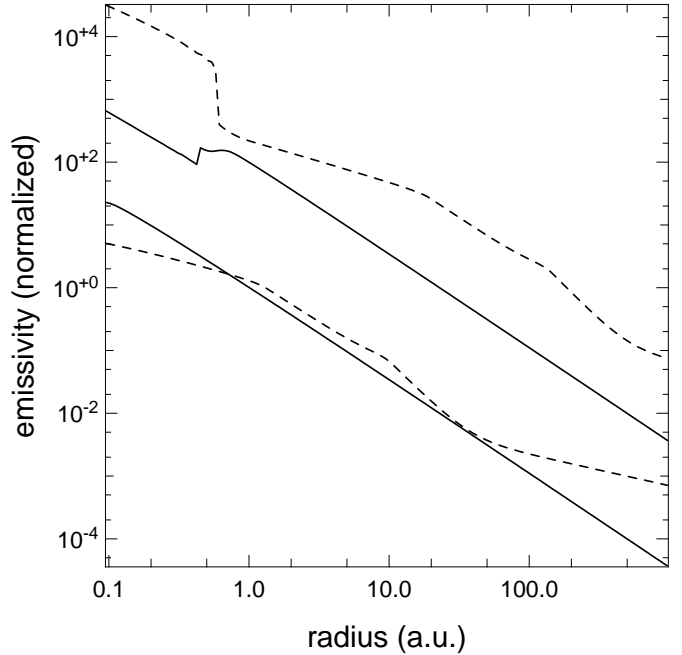}
  \caption{Emergent millimetric flux from a pole-on disk for several
    accretion rates and viscosity prescriptions.
    \textit{Upper curves:} $\mdot = 10^{-4}\,\msyr$;
    \textit{Lower curves:} $\mdot = 10^{-6}\,\msyr$;
    \textit{Solid lines:}  $\alpha=10^{-2}$; 
    \textit{Dashed lines:} $\beta=10^{-4}$.}
  \label{fig:emissivity}
\end{figure}

Radio observations can help us discriminate between the two viscosity
prescriptions, and especially radio observations in the outer disk, since they
are sensitive to the mass column.  The disk being optically thin in the
millimetric continuum, the flux is given by:
\begin{equation}
F_\nu \propto \int_{-\infty}^{+\infty} T(z) \rho(z) \,{\rm d}z \approx \Sigma \bar T 
\end{equation}
Figure~\ref{fig:emissivity} compares the millimetric emissivity for an
$\alpha$-disk and a $\beta$-disk with the same characteristics.  At
intermediate radii, the two emissivities are more or less equivalent,
however at larger radii the emissivities can follow somewhat different
power laws.  $F_\nu$ is proportionnal to $r^{-3/2}$ for an $\alpha$-disk
over the full range, whereas a $\beta$-disk presents a variable power-law
exponent.  In the domain where the disk is thermalized with the ISM,
\latin{i.e.} at large radii, $F_\nu \propto r^{-1/2}$.

\subsubsection{Interferometric observables}

\begin{figure*}[t]
   \parbox{0.48\hsize}{
      \null\hspace{0.02\hsize}\includegraphics[width=0.95\hsize]{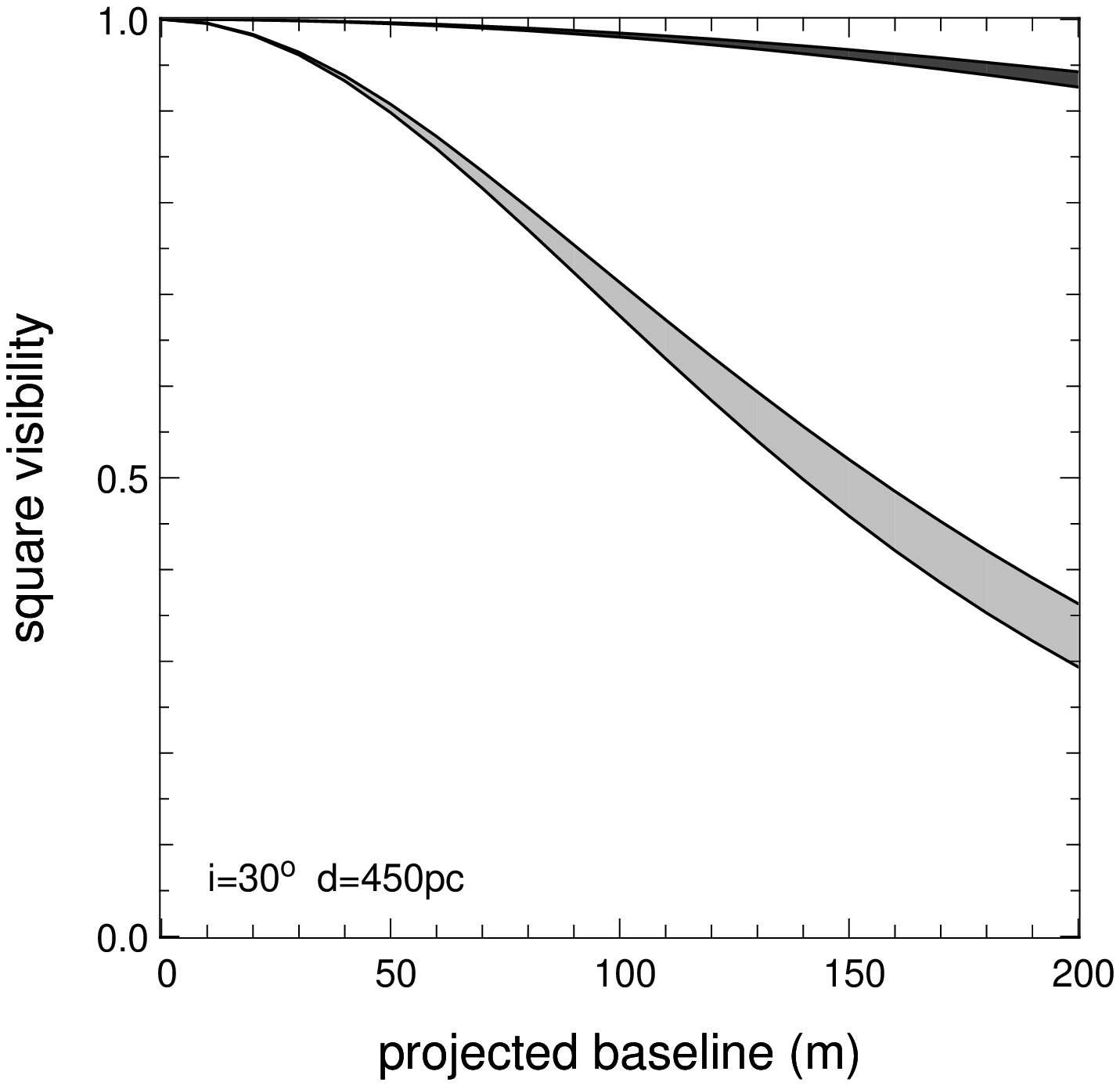}
      \caption{Visibilities from a disk model as a function of the baseline.
        \textit{Gray:} $\mdot = 10^{-4}\,\msyr$; \textit{Black:} $\mdot =
        10^{-6}\,\msyr$.  For a given baseline the visibility varies with
        the position angle; the domain between the minimum and maximum
        values is filled.}
      \label{fig:visib}
   }
   \hfill
   \parbox{0.48\hsize}{
      \null\hspace{0.02\hsize}\includegraphics[width=0.95\hsize]{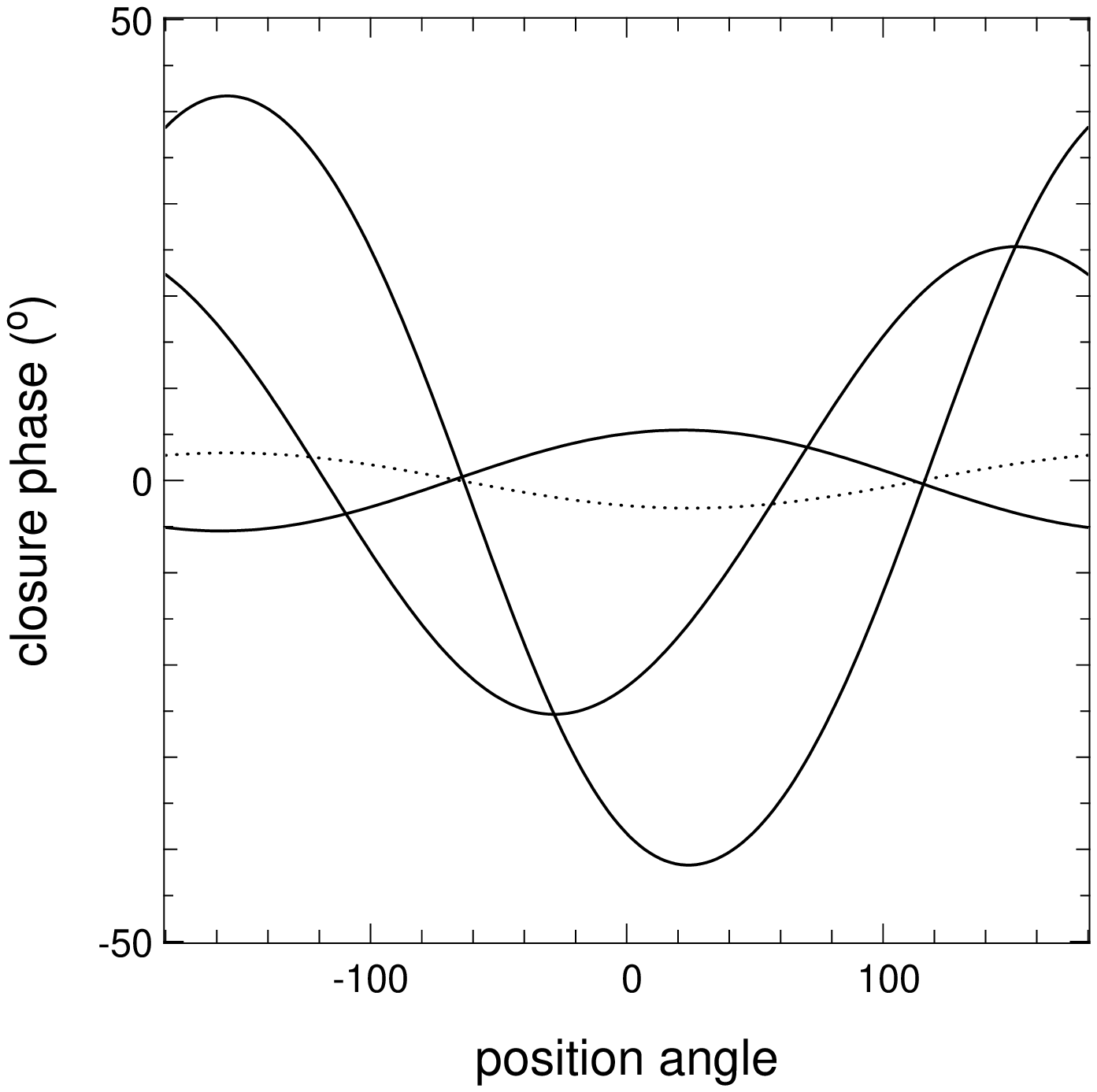}
      \caption{Closure phases as a function of the position angle for a
   triplet combination of the VLTI auxiliary telescopes. 
         \textit{Solid lines:} $\mdot = 10^{-4}\,\msyr$. 
         \textit{Dotted line:} $\mdot = 10^{-6}\,\msyr$ with the largest
         baseline.}   
      \label{fig:clph}
   }
\end{figure*}

Probing the inner parts of disks requires high angular resolution
techniques that can be achieved with present or soon-to-be operated
interferometric instruments \citep{mb95}.  The more massive disks ($\mdot >
10^{-6}\,\msyr$) should be resolved by the VLTI and its 100m-baseline, but
we should also detect non-zero closure phase.

Figure~\ref{fig:visib} shows the visibility of a disk as a function of the
projected baseline.  Since the image is not centro-symmetric, the visibility
depends on the position angle.  Figure~\ref{fig:clph} displays the closure
phase as a function of the position angle for different telescope combinations
in the VLTI for a disk with an important accretion rate ($\mdot =
10^{-4}\,\msyr$).  As a comparison, the closure phase is displayed for a disk
with moderate accretion rate with the widest combination of the VLTI auxiliary
telescopes. Because of the natural flaring, a disk that is not seen pole-on
does not appear centro-symmetric (as can be noticed in Fig.\ \ref{fig:images}),
therefore leading to a non-zero closure phase.

\subsubsection{Direct images}

\begin{figure*}[t]
\centerline{%
   \includegraphics[width=0.14\hsize]{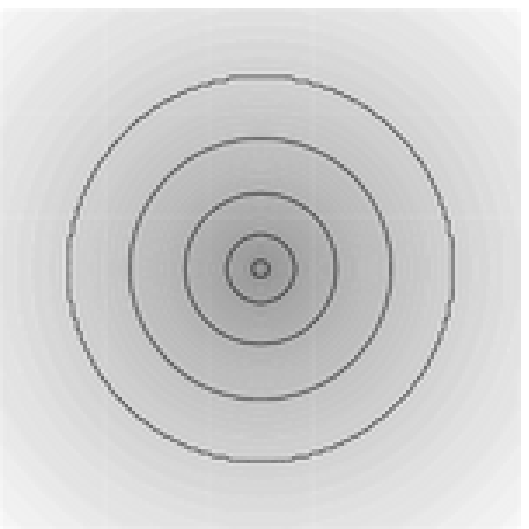}%
   \includegraphics[width=0.14\hsize]{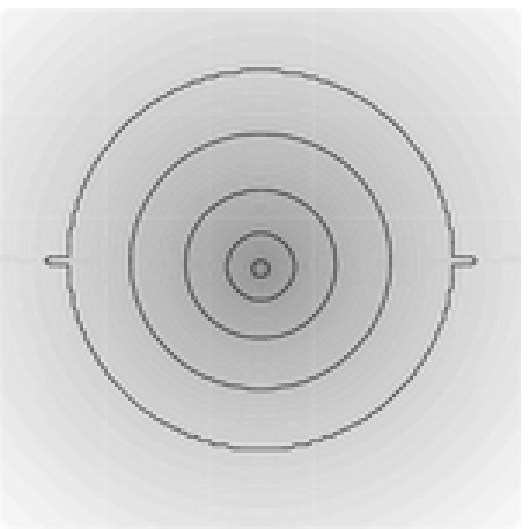}%
   \includegraphics[width=0.14\hsize]{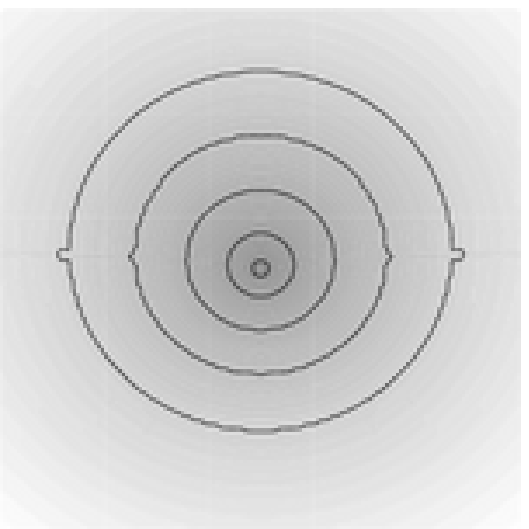}%
   \includegraphics[width=0.14\hsize]{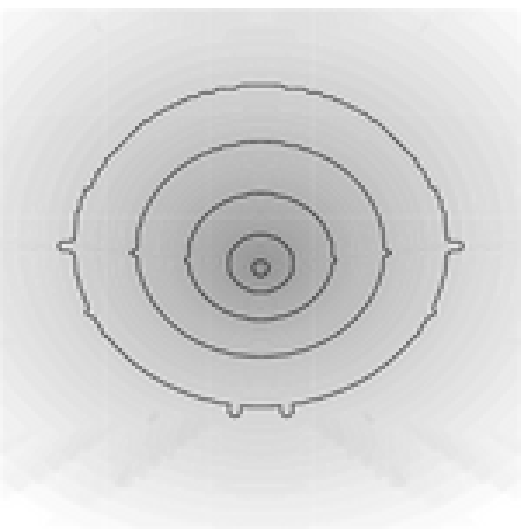}%
   \includegraphics[width=0.14\hsize]{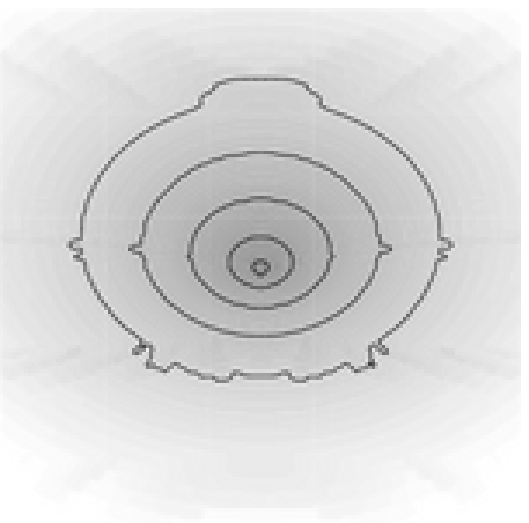}%
   \includegraphics[width=0.14\hsize]{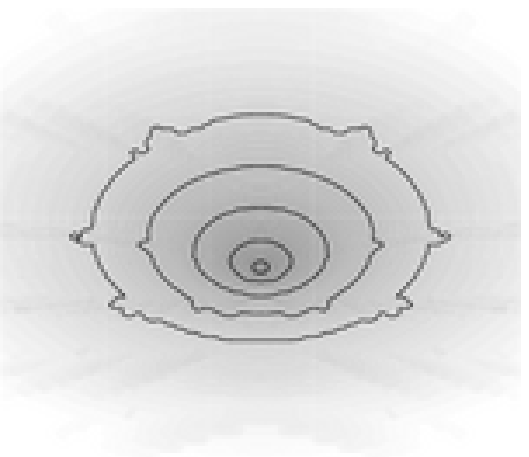}%
   \includegraphics[width=0.14\hsize]{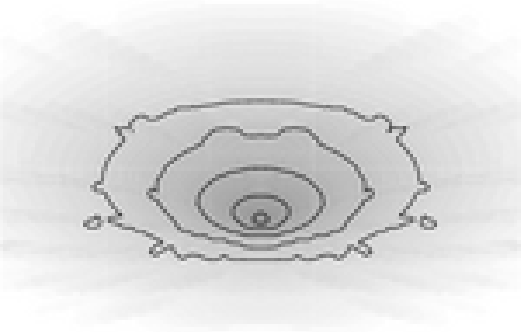}%
   }
   \caption{Synthetic images  of the thermal emission at 2.2\,$\mu$m for
     different inclinations of the disk (from $0^o$ to $60^o$ by steps of
     $10^o$). $\mdot = 10^{-6}\,\msyr$. The field is 50\,mas.}
   \label{fig:images}
\end{figure*}

Synthetic infrared images of the disk thermal emission are displayed in
Fig.~\ref{fig:images} with a field of 50 mas for a distance of $d =
450\,\mbox{pc}$.  As explained above, they are not centro-symmetric because of
the flaring.  So far, we have only access to interferometric observations at
this scale, so they are only relevant as a step to produce interferometric
observables.

\section{Conclusion}

We have presented a new method to model the vertical structure of accretion
disks. This method leads to analytical formulae for the temperature
distribution which help to understand the behaviour of the radiation propagated
inside the disk. Our model includes two sources of energy dissipation: viscous
heating and reprocessing of external radiation like that emitted by the
central star. We have applied these analytical results to the case of T Tauri
disks in a variety of conditions showing that the method is versatile: 
\begin{itemize}
\item with a modified Shakura-Sunyaev prescription for a disk with a
  non-constant accretion rate \citep{fer95}
   \item with a viscosity prescription following either the standard $\alpha$
   or the $\beta$ one proposed by \citet{hure00}
\end{itemize}

We are able to simulate the conditions of temperature and density in any part
of the circumstellar environment and to compute astronomical observables like
SED, optical and millimeter images or interferometric visibilities. This code
could be used also as starting conditions in a Monte-Carlo multiple scattering
code in order to derive polarization maps.

When completed, this code will offer to observers a tool based on physical
parameters to interpret their observations better than ad-hoc models based on
power laws often used today.


\begin{acknowledgements}
  We would like to thank E.~di Folco who has partly worked on the code. We
  are grateful to C.~Bertout, J.~Bouvier, C.~Ceccarelli, C.~Dougados and
  F.~Ménard for helpful discussions. We also thank the referee,
  Dr.~Hubeny, for useful suggestions.
\end{acknowledgements}

\appendix

\newcommand{\Teffk}{\ensuremath{(\Teff)_k}}
\newcommand{\thetak}{\ensuremath{\theta_k}}
\newcommand{\dtauhk}{\ensuremath{(\dtauh)_k}}
\newcommand{\uk}{\ensuremath{u_k}}
\newcommand{\Uk}{\ensuremath{U_k}}

\section{Vertical structure of the temperature with several energy sources}
\label{app:split}

Each source of energy locally contributes to the heating $\uk$ (see Sect.\ 
\ref{sec:model.diffsources}).  The total local heating is then $u = \sum_k
\uk$ and the energy dissipated per unit of disk surface is then $U = \sum_k
\Uk$ with
\begin{equation}
  \Uk M = \int_0^M \! \uk(\zeta) \diff{\zeta}.
\end{equation}
We now write the vertical distribution of energy dissipation, $\theta$ as a
function of the different vertical distributions of the energy source,
$\thetak$:
\begin{equation}
  \begin{split}
   \theta(m) &=        \int_0^m u  (\zeta) \diff\zeta / (UM)\\
             &= \sum_k \frac \Uk U \thetak(m) \label{eq:ap1}\\
\mbox{with}\quad \thetak(m) &= \int_0^m \uk  (\zeta) \diff\zeta / (\Uk M)\\
  \end{split}
\end{equation}

Similarly for $\dtauh$:
\begin{equation}
  \begin{split}
    \dtauh(m) &=        \int_0^m \theta (\zeta) \tauh(\zeta) \diff\zeta\\
              &= \sum_k \frac \Uk U \dtauhk \label{eq:ap2}\\
\mbox{with}\quad \dtauhk(m) &= \int_0^m \thetak (\zeta) \tauh(\zeta)
              \diff\zeta\\ 
  \end{split}
\end{equation}

Since,
\begin{align}
   \frac{u(m)}{U} &= \sum_k \frac \Uk U \,\, \frac{\uk(m)}{\Uk}
   \label{eq:ap4}
\end{align}
and   $\sum_k (\Uk/U) =1$, we can rewrite
Eq.~(\ref{eq:formsol1}) by replacing each term in the brackets by a sum
over the $k$ sources:
\begin{equation}
      \begin{split}
      T^4 = \frac{\kappaj \Teff^4}{4\kappab \fK} \,\,
      \sum_k \frac{\Uk}{U}\,
                \Bigg[
                   &\left( \tauh-\dtauhk + \frac{\fK(0)}{\fH} \right)\\
                   &+ \frac{\fK}{M \kappaj}\,\frac{\uk}{\Uk} 
                 \Bigg].
      \end{split}
   \label{eq:formsolk-0}
\end{equation}

If we define the effective temperature of the the $k$-th energy source, 
\begin{equation}
  \sigma_B (\Teff)_k^4 = \Uk M,
\end{equation}
then Eq.~(\ref{eq:formsol1}) can be written as the following sum:
\begin{equation}
      \begin{split}
      T^4 = \sum_k \frac{\kappaj (\Teff)_k^4}{4\kappab \fK}
                \Bigg[
                   &\left( \tauh-\dtauhk + \frac{\fK(0)}{\fH} \right)\\
                   &+ \frac{\fK}{M \kappaj}\,\frac{\uk}{\Uk} 
                 \Bigg]
      \end{split}
   \label{eq:formsolk-1}
\end{equation}
Therefore the vertical distribution of the temperature for a disk with
multiple energy dissipation sources is the same as the sum of the
individual temperature distributions using the global radiation in the
disk, \latin{i.e.} with $J$, $H$, $K$, $\tauh$ computed from the global
temperature distribution.

\section{Incoming radiation field}
\label{app:incrad}

We show in this appendix how to compute $\Jz(m)$, $\Js(m)$, $\Hz(m)$, and
$\Hs(m)$ in the case of an external point-like source.  The intensity of the
source is $\Jz$, giving us a boundary condition to solve the equations of
transfer (\ref{eq:transfer0}) and (\ref{eq:transfers}). If we define the
optical depths,
\begin{align}
   \tauz &= \int_0^m \chi^0_J(\zeta) \diff\zeta\\
   \taus &= \int_0^m \kappa^{\rm s}_J(\zeta) \diff\zeta
\end{align}
we obtain:
\begin{align}
   \Jz(m) &= J(0) {\rm e}^{-\tauz(m)/\muz} \label{eq:J0}\\
   \Js(m) &= \Lambda_{\taus(m)} \left( \frac{\sigmajz}{\kappajs} \Jz \right) 
             \label{eq:Js}
\end{align}
where $\Lambda$ is a commonly used operator  \citep[cf.][]{mih78} defined by:
\begin{align}
   &      &\Lambda_\tau \left(f\right) &= 
   \frac{1}{2} \int_0^{+\infty} E_1 \left| t - \tau \right| f(t) \diff{t}\\
   &{\rm and} &
   E_n(t) &= \int_1^{+\infty} u^{-n}{\rm e}^{-ut} \diff{u}.
\end{align}

The terms $\Hz(m)$ and $\Hs(m)$ can be computed in the same way as
$\Jz(m)$ and $\Js(m)$:
\begin{align}
  \Hz(m) &= \muz \, J(0) {\rm e}^{-\tauz(m)/\muz} \\
  \Hs(m) &=  \Phi_{\taus(m)} \left(\frac{\sigma^0_J}{\kappajs} \Jz \right)
\end{align}
with
\begin{align}                                %
  &&\Phi_\tau (f)   &=
         2 \int_0^{+\infty} E_2| t-\tau | f(t)
         \,\mathrm{sgn}(t-\tau) \diff{t}
\end{align}
where $\mathrm{sgn}(x)$ is the sign function.




\end{document}